\documentclass[twocolumn,superscriptaddress,amsmath,amssymb,aps,longbibliography,prx]{revtex4-2}
\usepackage[T1]{fontenc}
\usepackage{amssymb}
\usepackage{graphicx}
\usepackage{booktabs}
\usepackage{amsmath}
\usepackage{multirow}
\usepackage{blkarray}
\usepackage{nicematrix}
\usepackage{tikz}
\usepackage{mathrsfs}
\usepackage[bookmarks=true,colorlinks,citecolor=blue,urlcolor=blue]{hyperref}
\usepackage{braket}
\usepackage{babel}
\usepackage{blindtext}
\usepackage{soul}
\usepackage[compat=1.1.0]{tikz-feynman}
\usepackage[normalem]{ulem}
\usepackage{physics}

\usepackage{mathtools}
\usepackage{bbm}
\usepackage{tikz}
\usepackage{tabularx}
\usepackage{multirow}
\usepackage{graphicx}

\usepackage{amsthm}
\usepackage{txfonts}

\theoremstyle{definition}

\newcolumntype{C}[1]{>{\centering\arraybackslash}p{#1}}

\begin{document}

\def\papertitle{Continuous topological phase transition between $\mathbb{Z}_2$ topologically ordered phases}

\newcommand{\IOP}{\affiliation{Beijing National Laboratory for Condensed Matter Physics and Institute of Physics, Chinese Academy of Sciences, Beijing 100190, China}}
\newcommand{\TUM}{\affiliation{Technical University of Munich, TUM School of Natural Sciences, Physics Department, 85748 Garching, Germany}}
\newcommand{\MCQST}{\affiliation{Munich Center for Quantum Science and Technology (MCQST), Schellingstr. 4, 80799 M{\"u}nchen, Germany}}
\author{Qi Zhang}\IOP
\author{Wen-Tao Xu}\thanks{contact author: wen-tao.xu@tum.de}\TUM\MCQST

\date{\today}

\title{\papertitle}

\begin{abstract}
Topological phase transitions beyond anyon condensation remain poorly understood. A notable example is the transition between the toric code (TC) and double semion (DS) phases, which has two distinct $\mathbb{Z}_2$ topological orders in $(2+1)$D. Previous studies reveal that the transition between them can be either first order or via an intermediate phase, thus the existence of a direct continuous transition between them remains a long-standing problem.
Motivated by the fact that both phases can arise from  condensing distinct anyons in the $\mathbb{Z}_4$ topological order, we introduce a perturbed $\mathbb{Z}_4$ quantum double (QD) model to study the TC-DS transition. We confirm the existence of a continuous $(2+1)$D XY* transition between the TC and DS phases by mapping it to a two-coupled quantum Ising model.
Importantly, using the condensation order parameters and the area law coefficients of the Wilson loops, 
we further reveal that $\mathbb{Z}_4$ anyons, fractionalized from the $\mathbb{Z}_2$ topological orders, become deconfined at the transition between $\mathbb{Z}_2$ topologically ordered phases. 
Our results open a path toward developing a theoretical framework for topological phase transitions beyond anyon condensation.
\end{abstract}

\maketitle

\textbf{Introduction.}
Topological phases of matter, first discovered in the fractional quantum Hall effect~\cite{FQHE_1982,Laughlin_1983}, have revolutionized our understanding of quantum many-body systems.
Owing to their potential applications in quantum error correction and fault-tolerant quantum computing~\cite{Topo_quantum_memory_2002,kitaev_2002},
topological phases have become a central focus of both theoretical and experimental research.
Exactly solvable models with topologically ordered ground states have been constructed to study these phases~\cite{kitaev_2002, Kitaev_2006, String_net_2005}, and some of them have been experimentally realized using quantum processors and quantum simulators~\cite{TC_quantum_computer_2021,TC_Quantum_simulator_2021,D4_trapped_ion_2024,doublefib:2024}. 
As phases of matter beyond the conventional Ginzburg-Landau-Wilson (GLW) paradigm of spontaneous symmetry breaking, topologically ordered states pose significant challenges for studying their phase transitions within the framework of the GLW paradigm.
Nonetheless, topological phase transitions driven by anyon condensation can be partially understood using the GLW paradigm, because they can be mapped, via duality transformations, to spontaneous symmetry-breaking phase transitions characterized by local order parameters~\cite{Trebst_2007,Zhu_2019,Adam_Nahum_2021,Xu_2021,Xu_2022,TC_hyc_tri_lat_2024}. Such topological phase transitions have been extensively studied using various models and methods~\cite{MCP_TC_2010, Youjin_2012, haegeman2015shadows, Iqbal_2017, cond_deriven_2017, Iqbal_2018, Xu_2020, Qi_Zhang_2020, Kao_2021, Iqbal_Schuch_2021, Xu_2021, Xu_2022, TFD_state_2024,TC_hyc_tri_lat_2025}.

\begin{figure}
    \centering
    \includegraphics[width=1\linewidth]{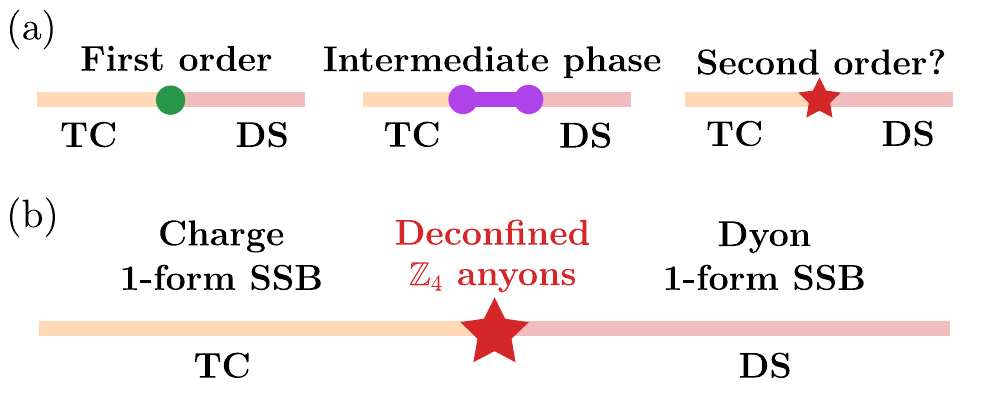}
    
    \caption{\textbf{Schematic illustration of the motivation and main results.} (a) Previous works~\cite{TC_DS_ED,TC_DS_QMC_2021,Scaffidi_2023} reveal that the transition between the TC and DS phases can be first order or via an intermediate phase. The existence of a direct continuous transition between them is an open question. (b) We confirm a direct continuous TC-DS transition point, where the adjacent TC and DS phases spontaneously break two different 1-form symmetries. We further reveal deconfined $\mathbb{Z}_4$ anyons at their transition point.
    }
    \label{fig:tc2ds}
\end{figure}

However, topological phase transitions that are \emph{not} induced by anyon condensation remain poorly understood. A notable example is the transition between the toric code (TC)~\cite{kitaev_2002} and double semion (DS) phases~\cite{String_net_2005}, which represent two distinct $\mathbb{Z}_2$ topological orders with incompatible anyons.
By interpolating the fixed-point TC and DS string-net models using a one-parameter Hamiltonian, the transition has been studied using exact diagonalization~\cite{TC_DS_ED} and quantum Monte Carlo~\cite{TC_DS_QMC_2021,Scaffidi_2023}, the former method indicates a first-order transition, while the latter method suggests the existence of an intermediate gapless phase, as shown in Fig.~\ref{fig:tc2ds}a.
Additionally, a fine-tuned transition between the TC and DS phases can be realized using deformed projected entangled pair states (PEPS) with suitable tuning parameters~\cite{Iqbal_2017,Iqbal_2018,Xu_zhang_2018,Leo_2024}. Furthermore, a field-theoretic argument has been proposed for a continuous transition between the TC and DS phases in the presence of SU(2) symmetry~\cite{CSL_Z2_2013}.
Despite these efforts, how to concretely realize a direct continuous transition between the TC and DS phases using a generic Hamiltonian remains a long-standing open question. 
If such a continuous transition exists, the nature of the critical point, particularly how the incompatible anyons of the two $\mathbb{Z}_2$ topological orders reconcile at the critical point, remains unclear.

In this letter, we confirm the existence of a direct continuous TC-DS transition and provide new insights into its nature. Rather than considering a model interpolating directly between fixed points of the TC and DS phases~\cite{TC_DS_ED,TC_DS_QMC_2021,Scaffidi_2023}, we propose a perturbed \(\mathbb{Z}_4\) quantum double (QD) model based on the observation that both phases can emerge from the \(\mathbb{Z}_4\) topological order via anyon condensation~\cite{Iqbal_2017,Iqbal_2018,Dominic_2022}. 
To determine its phase diagram and the quantum criticality of the continuous TC-DS transition, we introduce a duality transformation that maps the perturbed $\mathbb{Z}_4$ QD model to a two-coupled $(2+1)$D quantum Ising model, and approximate its ground states using variationally optimized infinite PEPS (iPEPS)~\cite{iPEPS_corboz_2016,iPEPS_Laurens_2016} with automatic differentiation~\cite{AD_2019}.

Using the condensation order parameters and the area law coefficients of the Wilson loop expectation values, we identify that the continuous TC-DS transition belongs to the XY* universality class. Interestingly, as shown in Fig.~\ref{fig:tc2ds}b, the continuous TC-DS transition resembles a deconfined quantum critical point~\cite{DQCP_2004,DQCP_PRB_2004}—a direct continuous transition between two phases that spontaneously break different symmetries and exhibit fractionalized excitations—in the following aspects: First, the perturbed $\mathbb{Z}_4$ QD model has two distinct 1-form symmetries, which are loop-like generalized symmetries. One 1-form symmetry is broken spontaneously in the TC phase, while the other one is broken spontaneously in the DS phase.
Second, some anyons in the $\mathbb{Z}_2$ topological orders are fractionalized and become deconfined at the continuous TC-DS transition point. 
Our results suggest that, at the critical point between topological phases that cannot be connected via anyon condensation, fractionalized anyons that are confined in both adjacent phases can emerge. This opens a path toward developing a general theoretical framework for topological phase transitions beyond anyon condensation.

\begin{figure}
    \centering
    \includegraphics[width=1\linewidth]{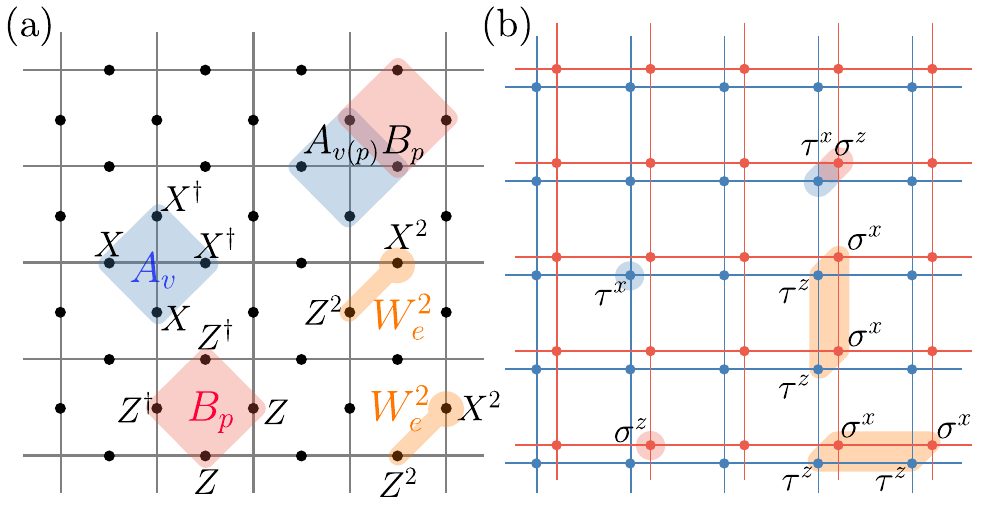}
    \caption{\textbf{Perturbed $\mathbb{Z}_4$ QD model and duality transformation}. (a) The multi-site local Hamiltonian terms of the perturbed $\mathbb{Z}_4$ QD model, $e$ of $W_e^2$ is the edge where the $X^2$ operator is located. (b) The perturbed $\mathbb{Z}_4$ QD model is mapped to a two-coupled Ising model, the blue (red) lattice is the primal (dual) lattice, and we shift the dual lattice to overlap with the primal lattice.
    }
    \label{fig:model}
\end{figure}

\textbf{A perturbed $\mathbb{Z}_4$ quantum double model.} 
Because the model interpolating between the fixed-point Hamiltonians of the TC and DS phases~\cite{TC_DS_ED,TC_DS_QMC_2021} is difficult to analyze analytically and simulate numerically, we alternatively propose a perturbed $\mathbb{Z}_4$ QD model to realize the topological phase transition between the TC and DS phases~\footnote{The $\mathbb{Z}_4$ QD model is also known called \(\mathbb{Z}_4\) toric code model. In this work, \textquotedblleft TC\textquotedblright{} refers exclusively to the \(\mathbb{Z}_2\) toric code model, and we avoid calling \(\mathcal{H}_{\text{QD}}\) the \(\mathbb{Z}_4\) TC model to prevent confusion.}. 
The fixed-point Hamiltonian for the \(\mathbb{Z}_4\) QD phase
on the square lattice is~\cite{kitaev_2002}: 
\begin{equation}
   \mathcal{H}_{\text{QD}} = -\left(\sum_v A_v + \sum_p B_p + \text{h.c.} \right), 
\end{equation}
where the vertex and plaquette operators, \(A_v\) and \(B_p\), are illustrated in Fig.~\ref{fig:model}a, and are defined using the \(\mathbb{Z}_4\) qudit operators: $ X = \sum_{n=0}^{3} \ket{n}\bra{(n+1) \bmod 4},  
    Z = \sum_{n=0}^{3} \mathrm{i}^n \ket{n}\bra{n}$.
Electric charges \(\{\pmb{e}, \pmb{e}^2, \pmb{e}^3\}\) correspond to the eigenvalues \(\{\mathrm{i}, -1, -\mathrm{i}\}\) of \(A_v\), while magnetic fluxes \(\{\pmb{m}, \pmb{m}^2, \pmb{m}^3\}\) correspond to the same eigenvalues of \(B_p\). 
Their composites, \(\pmb{e}^n \pmb{m}^k\) ($n,k\in\{1,2,3\}$), are known as dyons. 
Notice that charges (fluxes) are located on the primal (dual) lattice, and we adopt the convention in Ref.~\cite{Dominic_2022} that dyons live on the dual lattice. 
Further details on the anyon content of the $\mathbb{Z}_2$ topological order are provided in the Supplemental Material (SM)~\cite{appendix}.

According to anyon condensation theory~\cite{Anyon_cond_theory_2002,Anyon_cond_theory_2009,Liang_kong_2014,anyon_cond_2018}, the \(\mathbb{Z}_4\) QD model can undergo phase transitions to the TC phase by condensing the charge \(\pmb{e}^2\)~\footnote{Alternative, the phase transitions to the TC phase by condensing the flux \(\pmb{m}^2\) by considering the Hamiltonian \(\mathcal{H}_{\mathrm{QD}} - h_x \sum_e X_e^2\) with $h_x>0.657$. }, and to the DS phase by condensing the dyon \(\pmb{e}^2\pmb{m}^2\)~\cite{Iqbal_2017,Iqbal_2018}. 
Microscopically, the transition to the TC phase can be realized by a Hamiltonian \(\mathcal{H}_{\mathrm{QD}} - h_z \sum_e Z_e^2\) with $h_z\gtrsim0.657$
~\footnote{The phase transition point can be determined by mapping the Hamiltonian \( \mathcal{H}_{\mathrm{QD}} - h_z \sum_e Z_e^2 \) to the transverse-field Ising model \( H_{\text{Ising}} = -\sum_{\langle i,j\rangle} \sigma^z_i \sigma^z_j - h \sum_i \sigma^x_i \). This yields the critical point at \( h_z = 2 / h_c^{\text{TFIM}} \), where \( h_c^{\text{TFIM}} = 3.044\,330(6) \) is the known critical value of the transverse-field Ising model~\cite{youjin_2020}.}.
In contrast, dyon condensation is less straightforward. Interestingly, the fixed-point DS model can be constructed using \(\mathbb{Z}_4\) qudit operators~\cite{Dominic_2022}:
\begin{equation}
    \mathcal{H}_{\text{DS}} = -\sum_{p} \left(A_{v(p)} B_p + A_{v(p)} B_p^{\dagger} + \text{h.c.}\right) - \sum_e W^2_e,
\end{equation}
where $v(p)$ denotes the vertex in the southwest of and adjacent to the plaquette $p$ (see Fig.~\ref{fig:model}a). The first term penalizes \emph{semionic} dyon excitations \(\pmb{e}^n\pmb{m}^k\) with odd \(n,k\)~\footnote{Note that \(A_v B_p A_v B_p^{\dagger} = A_v^2\) and \(A_v B_p A_v^{\dagger} B_p = B_p^2\) are conserved quantities. Although the Hamiltonian used here differs from \(\mathcal{H}'_{\text{DS}} = -\sum_{\langle vp \rangle'} (A_v B_p + B_p^2 + \text{h.c.}) - \sum_e W_e\) in Ref.~\cite{Dominic_2022}, they share the same ground state.}. 
The last term $W^2_e=X^2_eZ^2_{e'}$ induces the condensation of the \emph{bosonic} dyon \(\pmb{e}^2\pmb{m}^2\), where $e'$ denotes the nearest neighboring edge in the southwest of the edge $e$.

By combining \(\mathcal{H}_{\text{QD}}\), \(\mathcal{H}_{\text{DS}}\), and the perturbation terms \(-\sum_e X_e^2\) and \(-\sum_e Z_e^2\), we propose the following perturbed $\mathbb{Z}_4$ QD model to study topological phase transitions between the TC and DS phases:
\begin{align}\label{eq:Z4_QD_H}
    H_{\text{QD}} &= -\left(\sum_v A_v + \sum_p B_p + \text{h.c.}\right)
    - h_x \sum_e X_e^2 - h_z \sum_e Z_e^2 \notag\\
    &\quad - \frac{1}{2} \sum_{p} \left(A_{v(p)} B_p + A_{v(p)} B_p^{\dagger} + \text{h.c.}\right)
    - h_w \sum_e W^2_e.
\end{align}
The Hamiltonian possesses an extensive set of conserved quantities: $[A^2_v, H_{\text{QD}}] = 0$ for all vertices, and $[B^2_p, H_{\text{QD}}] = 0$ for all plaquettes.
Products of these operators generate the so-called 1-form symmetries~\cite{High_form_Kapistin_2015,McGreevy_2023,Bhardwaj:2023}, represented by the non-local loop operators $\prod_{e\in C}Z_e^2$, $\prod_{e\in\hat{C}}W_e^2$, $\prod_{e\in\hat{C}}X_e^2$, 
where $C$ and $\hat{C}$ denote loops on the original and dual lattices, respectively. Since different topological orders can be understood as distinct patterns of higher-form symmetry breaking~\cite{High_form_wen_2019}, 
the 1-form symmetries can be used to characterize different topological phases.

\textbf{Condensation order parameters and area law coefficients.}
We use non-local order parameters to detect spontaneous 1-form symmetry breaking, which distinguishes different topological phases. 
One class of such non-local order parameters includes
$\left\langle\prod_{e\in S}Z_e^2\right\rangle$, $\left\langle\prod_{e\in\hat{S}}W_e^2\right\rangle$, and $\left\langle\prod_{e\in\hat{S}}X_e^2\right\rangle$~\cite{BRICMONT_1983,xu_FM_2024}, where $S$ ($\hat{S}$) denotes the open strings on the original (dual) lattices.
They can be viewed as segments of closed loops $C$ ($\hat{C}$), as detailed in SM~\cite{appendix}.
Because the anyons $\pmb{e}^2$, $\pmb{e}^2\pmb{m}^2$ and $\pmb{m}^2$ are created at the endpoints of the string operators, their non-zero expectation values imply the corresponding anyons condense, as summarized in Tab.~\ref{tab:order_para}. We therefore call them condensation order parameters.

\begin{table}[]
    \centering
    \caption{\textbf{Condensation order parameters and area law coefficients.} $S$ ($\hat{S}$) denotes a semi-infinite long string on the primal (dual) lattice, $C$ ($\hat{C}$) denote the a loop on the primal (dual) lattice, $R$ denotes the set of sites surround by the loop $C$ ($\hat{C}$) and $a_X$ is the area of the shape $X$.
    The column ``Def. in $H_{\text{QD}}$'' lists the definition of the condensation order parameters and area law coefficients in the $\mathbb{Z}_4$ QD model, while ``Def. in $H_{\text{dual}}$'' for the two-coupled Ising model.
    The columns ``$\mathbb{Z}_4$ QD'', ``DS'', ``TC'' and ``Trivial'' show values of these quantities in the respective phases.
    }
    \begin{tabular}{p{0.20cm}C{0.9cm} C{1.9cm} C{1.6cm} C{0.9cm}C{0.6cm}C{0.6cm}C{0.8cm}}
    \toprule\toprule
   & Label & Def. in $H_{\text{QD}}$ & Def. in $H_{\text{dual}}$ & $\mathbb{Z}_4$ QD& DS & TC & Trivial \\
    \midrule
    \multirow{3}{*}{\rotatebox{90}{Cond.~~~~}}&$\langle\pmb{e}^2\rangle$ & $\left|\left\langle \prod_{e\in S}Z^2_e\right\rangle\right|^{1/2}$ & $\left|\left\langle \tau_i^z\right\rangle\right|$ &0 & 0  &$>0$ &  $>0$\\
    &$\langle\pmb{e}^2\pmb{m}^2\rangle$ & $\left|\left\langle \prod_{e\in \hat{S}}W^2_e\right\rangle\right|^{1/2}$ & $\left|\left\langle \tau_i^z\sigma_i^x\right\rangle\right|$ & 0 & $>0$&$0$ &  $>0$\\
    &$\langle\pmb{m}^2\rangle$ & $\left|\left\langle \prod_{e\in \hat{S}}X^2_e\right\rangle\right|^{1/2}$ & $\left|\left\langle \sigma_i^x\right\rangle\right|$ &0 & 0&$0$ &  $>0$ \\
    \midrule
    \multirow{3}{*}{\rotatebox{90}{Conf.~~~~~~}}
    &$\alpha(\pmb{e})$ & $\frac{\ln\left\langle \prod_{e\in C}Z^{\gamma(e)}_e\right\rangle}{-a_{C}}$ & $\frac{\ln\langle\prod_{i\in R} \sigma_i^z\rangle}{-a_{R}}$ & 0 & $>0$&$0$ &  $>0$ \\
    &$\alpha(\pmb{e}\pmb{m})$ & $\frac{\ln\left\langle\prod_{e\in \hat{C}}X^{\hat{\gamma}(e)}_eZ^{\gamma(e')}_{e'}\right\rangle}{-a_{\hat{C}}}$ & $\frac{\ln\langle\prod_{i\in R} \tau^x_i\sigma^z_i\rangle}{-a_{R}}$ & 0 & 0&$>0$ &  $>0$ \\
    &$\alpha(\pmb{m})$ & $\frac{\ln\left\langle \prod_{e\in \hat{C}}X^{\gamma(e)}_e\right\rangle}{-a_{\hat{C}}}$ & $\frac{\ln\langle\prod_{i\in R} \tau^x_i\rangle}{-a_{R}}$ & 0 & $>0$&$>0$ &  $>0$ \\
    \bottomrule
    \bottomrule
    \end{tabular}
    \label{tab:order_para}
\end{table}

Another class of non-local order parameters is defined using Wilson loop operators $\left\langle\prod_{e\in C}Z^{\gamma(e)}_e\right\rangle$, 
$\left\langle\prod_{e\in \hat{C}}X^{\hat{\gamma}(e)}_eZ^{\gamma(e')}_{e'}\right\rangle$ and 
$\left\langle\prod_{e\in \hat{C}}X^{\hat{\gamma}(e)}_e\right\rangle$, 
where $e'$ denotes the edge immediately southwest of $e$,
and the exponents $\gamma(e),~\hat{\gamma}(e)$, taking values $1$ or $3$, depend on the edge orientations~\cite{Dominic_2022, appendix}. It is well known that the expectation values of the Wilson loop operators can detect the confinement of anyons~\cite{Wegner_duality_1971,Kogut_1979}. 
Let us focus on $\left\langle\prod_{e\in C}Z^{\gamma(e)}_e\right\rangle$, where the loop operator creates a pair of anyons $\pmb{e}$ and $\pmb{e}^3$ and then annihilates them. 
The expectation value satisfies $\left\langle\prod_{e\in C}Z^{\gamma(e)}_e\right\rangle\propto\mathrm{e}^{-\alpha(\pmb{e}) a_C}$ for a large loop $C$ that surrounds the area $a_C$, where $\alpha(\pmb{e})$ is the area law coefficient. 
When $\alpha(\pmb{e})>0$, it is the area law, indicating that $\pmb{e}$ is confined.
If $\alpha(\pmb{e})=0$, it implies that $\left\langle\prod_{e\in C}Z^{\gamma(e)}_e\right\rangle$ decays at most as the perimeter of the loop $C$, i.e. the perimeter law, indicating that $\pmb{e}$ is deconfined. 
Therefore, we can use the area law coefficient $\alpha(\pmb{e})$ to distinguish the deconfinement and confinement of $\pmb{e}$. 
A similar argument applies to $\left\langle\prod_{e\in \hat{C}}X^{\hat{\gamma}(e)}_e Z^{\gamma(e')}_{e'}\right\rangle$ and $\left\langle\prod_{e\in \hat{C}}X^{\gamma(e)}_e\right\rangle$, which is summarized in Tab.~\ref{tab:order_para}. 

\begin{figure}[h]
    \centering
    \includegraphics[width=0.7\linewidth]{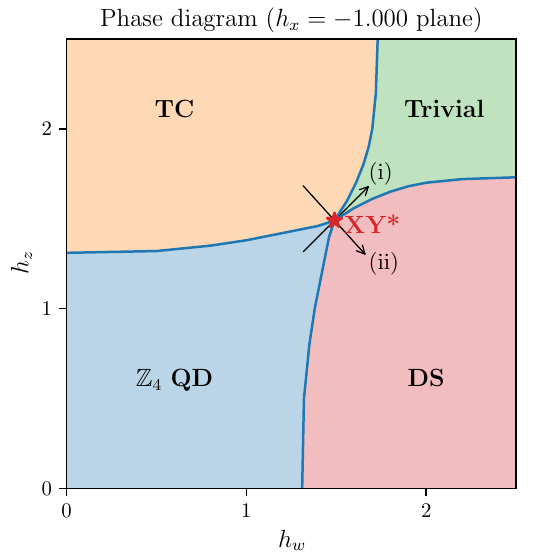}
    \caption{ 
    \textbf{Phase diagram of the perturbed $\mathbb{Z}_4$ QD model on the $h_x=-1$ plane.} The model has three tuning parameters: $h_x$, $h_z$ and $h_w$. The solid blue lines indicate the 3D Ising* transitions. The red star marks the multi-critical point, which belongs to the XY* universality class. Arrows (i) and (ii) denote the lines $h_z=h_w$ and $h_w=2.980-h_z$, respectively, both crossing the multi-critical point.
    }
    \label{fig:phasediagram}
\end{figure}

\textbf{Mapping to a two-coupled (2+1)D quantum Ising model.} 
The 1-form symmetries not only characterize different topological phases but also simplify the perturbed $\mathbb{Z}_4$ QD model in Eq.~\eqref{eq:QAT_H} via a duality transformation.
It is well known that the $\mathbb{Z}_2$ lattice gauge theory with the exact 1-form Wilson or 't Hooft loop symmetry can be mapped to the transverse-field Ising model~\cite{youjin_2020} via the Wegner duality transformation~\cite{Wegner_duality_1971, Trebst_2007}.
In the same spirit, we propose the following duality transformation for the perturbed $\mathbb{Z}_4$ QD model:
\begin{equation}\label{duality_transformation}
    A_v+A_v^{\dagger}\rightarrow 2\tau_v^x,\,\, B_p+B_p^{\dagger}\rightarrow 2\sigma_p^z,\,\, X^2_e\rightarrow \sigma^x_{p}\sigma^x_{p'},\,\,Z^2_e\rightarrow \tau^z_{v}\tau^z_{v'},\notag
\end{equation}
where $\{\sigma_p^{\alpha}\}$ ($\{\tau_v^{\alpha}\}$) with $\alpha\in\{x,y,z\}$ is a set of Pauli matrices of $\mathbb{Z}_2$ qubits applying on the site of the dual (primal) lattice.
As shown in Fig.~\ref{fig:model}b,
by shifting the dual lattice such that it completely overlaps the primal lattice, and relabeling $v,v'$ and $p,p'$ using $i,j$, the perturbed $\mathbb{Z}_4$ QD model in Eq.~\eqref{eq:Z4_QD_H} is mapped to a two-coupled $(2+1)$D quantum Ising model:
\begin{align}\label{eq:QAT_H}
    H_{\text{dual}}=&-2\sum_i \left(\tau^x_i+ \sigma_i^z+\tau^x_i\sigma^z_i\right)-h_x\sum_{\langle ij\rangle} \sigma^x_i\sigma^x_{j}-h_z\sum_{\langle ij\rangle} \tau^z_i\tau_{j}^z \notag\\
    &-h_w\sum_{\langle ij\rangle} \tau^z_i\tau_{j}^z\sigma^x_i\sigma^x_{j}.
\end{align}
When taking the boundary condition into account, i.e., consider the system on a torus, the duality mapping becomes $
    H_{\text{QD}}=\bigoplus_{b^{\sigma}_x,b^{\sigma}_y,b^{\tau}_x,b^{\tau}_y} \frac{1+\prod_i \sigma^z_i}{2}\frac{1+\prod_i \tau^x_i}{2}H_{\text{dual}}(b^{\sigma}_x,b^{\sigma}_y,b^{\tau}_x,b^{\tau}_y),
$
where $b^{\sigma}_x,b^{\sigma}_y,b^{\tau}_x,b^{\tau}_y$ are boundary conditions in $x$ and $y$ directions of the $\sigma$ or $\tau$ spins which can be either periodic ($+$) or antiperiodic ($-$), and the anti-periodic boundary condition means that the ferromagnetic couplings along some non-contractible loops are changed to the anti-ferromagnetic couplings. In the SM~\cite{appendix}, we further analyze that the symmetry sectors and boundary conditions induced by the duality transformation do not affect the evaluation of order parameters.
If the term $\sum_i \tau_i^x\sigma^z_i$ is absent and $h_x=h_z$, it becomes a standard $(2+1)$D quantum Ashkin-Teller model. 
It is obvious that the phase diagram of $H_{\text{dual}}$ is symmetric about $h_x=h_z$, because the model is invariant under the exchange of $\sigma_i^x\leftrightarrow\tau_i^z$, $\sigma_i^z\leftrightarrow\tau_i^x$, and $h_x\leftrightarrow h_z$. Furthermore, its phase diagram is also symmetric about $h_z=h_w$, because $UH_{\text{dual}}(h_x,h_z,h_w)U^{\dagger}=H_{\text{dual}}(h_x,h_w,h_z)$, where the unitary transformation is $U=\prod_i \left(\frac{1+\sigma_i^x}{2}+\tau_i^x\frac{1-\sigma_i^x}{2}\right)$. 
Since duality transformations preserve the structure of phase diagrams,
we can directly deduce the phase diagram of the $\mathbb{Z}_4$ QD model $H_{\text{QD}}$ from that of $H_{\text{dual}}$.

The non-local condensation order parameters and the Wilson loop expectation values of the perturbed $\mathbb{Z}_4$ QD model can be mapped to the local order parameters and the membrane-like disorder parameters of the two-coupled quantum Ising model~\cite{Kadanoff_1971,Fradkin2017} (see SM~\cite{appendix} and Tab.~\ref{tab:order_para}). 
This allows us to extract condensation order parameters and the area law coefficients directly from the ground states of the two-coupled Ising model, which can be approximated using $C_{4v}$-symmetric iPEPS~\cite{iPEPS_Laurens_2016,iPEPS_corboz_2016,AD_2019}.
Compared to other methods such as quantum Monte Carlo, iPEPS is more suitable to evaluate non-local order parameters, and the method of extracting area law coefficients from iPEPS is proposed in Ref.~\cite{xu_huang_2025}. By comparing the numerical results of the condensation order parameters and the area law coefficients with Tab.~\ref{tab:order_para}, 
we obtain the phase diagrams for various $h_x$ planes (see SM~\cite{appendix}), one of which is shown in Fig.~\ref{fig:phasediagram}.
It includes the $\mathbb{Z}_4$ QD, TC, DS, and trivial phases, and their phase boundaries are described by the 3D Ising* universality class describing 1-form symmetry-breaking transitions~\cite{Universal_CFT_spec_2016}. 
Next, we focus on the multi-critical point, especially the continuous TC-DS transition.

\begin{figure*}[tbp]
    \centering
    \includegraphics[width=1\linewidth]{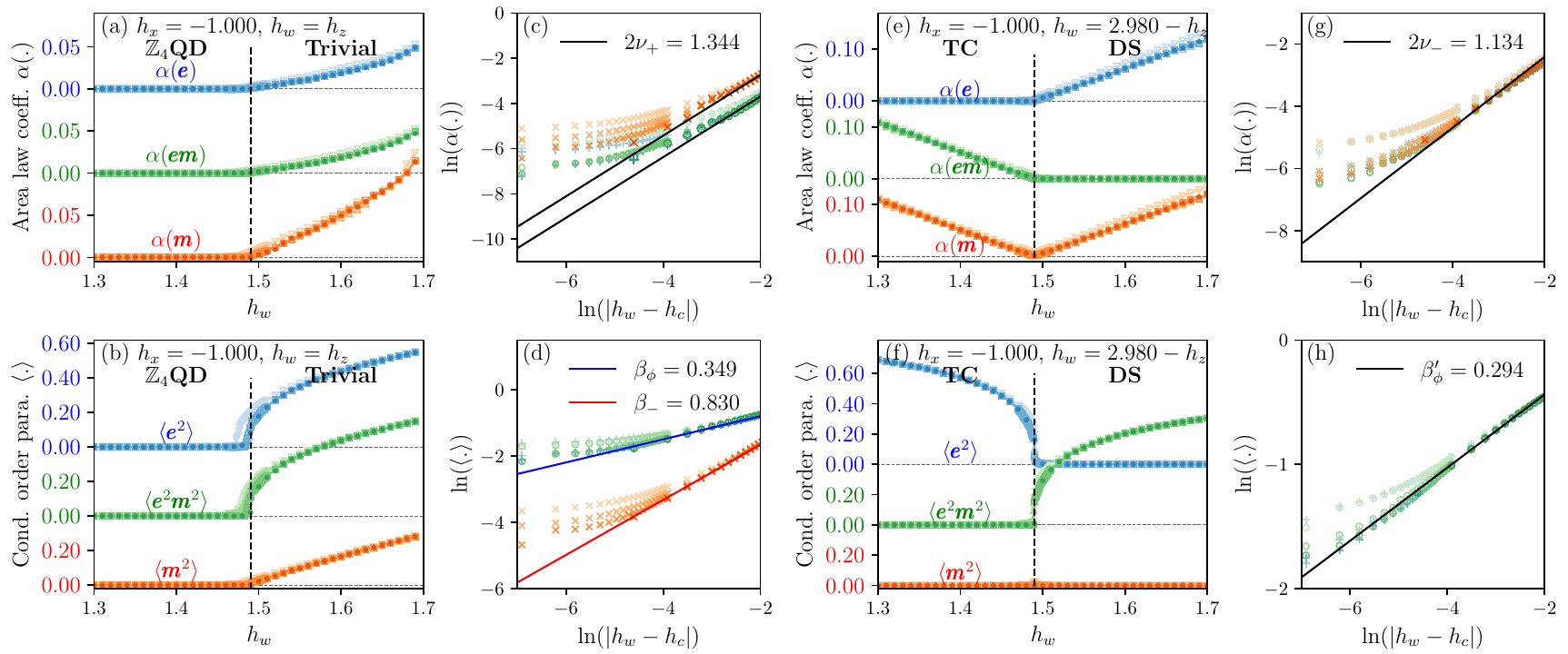}
    \caption{\textbf{Condensation order parameters and area law coefficients near and at the continuous TC-DS transition point $(h_x,h_z,h_w)\approx(-1.000,1.490,1.490)$. } Different colors correspond to different order parameters, and the colors from light to dark correspond to the bond dimensions $(D,\chi)=(4,100),(5,100),(6,100),(7,100)$, where $D$ is the iPEPS bond dimension and $\chi$ is the bond dimension of the corner transfer matrix renormalization group method contracting the iPEPS~\cite{CTMRG_1,CTMRG_2,Fast_CTMRG}. The vertical dashed line marks the location of the multi-critical point ($h_c=1.490$). From the XY* universality class, we have three order parameter critical exponents $\beta_{\phi}$,  $\beta_{\phi}'$,  $\beta_{-}$ and two correlation length critical exponents $\nu_{+}$ and $\nu_{-}$.
    (a) Area law coefficients along the line (i) in Fig.~\ref{fig:phasediagram}.
    (b) Condensation order parameters along the line (i).
    (c) Double-log plot of the area law coefficients in (a), where the slope of the straight lines is $2\nu_{+}$. 
    (d) Double-log plot of the condensation order parameters in (b), where the slopes of the straight lines are $\beta_{\phi}$ and $\beta_{-}$, respectively.
    (e) Area law coefficients along the line (ii) in Fig.~\ref{fig:phasediagram}.
    (f) Condensation order parameters along the line (ii). 
    (g) Double-log plot of the area law coefficients in (e), where the slope of the straight lines is $2\nu_{-}$. 
    (h) Double-log plot of the condensation order parameters in (f), where the slope of the straight lines is $\beta_{\phi}'$.}
    \label{fig:xy}
\end{figure*}

\textbf{Continuous topological phase transition between the $\mathbb{Z}_4$ QD phase and the trivial phase.}
By tuning $h_x$ to negative values, we find that a continuous TC-DS phase transition becomes possible.
We focus on the $h_x=-1$ plane (Fig.~\ref{fig:phasediagram}), where a continuous TC-DS transition emerges at the junction of four phases: the $\mathbb{Z}_4$ QD, TC, DS, and trivial phases. 
Since the phase diagram is symmetric about $h_z=h_w$ [line (i) in Fig.~\ref{fig:phasediagram}] and the continuous TC-DS transition must occur at a point on this line, we first analyze the condensation order parameters and area law coefficients along this line.

Fig.~\ref{fig:xy}a shows that the area law coefficients $\alpha(\pmb{e})$, $\alpha(\pmb{e}\pmb{m})$ and $\alpha(\pmb{m})$ are zero when $h_w=h_z<h_c\approx 1.490$, but they are non-zero when $h_w=h_z>h_c$, implying $\pmb{e}^n$, $\pmb{e}^n\pmb{m}^k$ and $\pmb{m}^k$ (\(n,k\in\{1,3\}\)) are confined. Moreover, Fig.~\ref{fig:xy}b shows that the condensation order parameters $\langle \pmb{e}^2\rangle$, $\langle \pmb{e}^2\pmb{m}^2\rangle$ and $\langle \pmb{m}^2\rangle$ are zero when $h_w=h_z<h_c$, but they are non-zero when $h_w=h_z>h_c$, implying $\pmb{e}^2$, $\pmb{e}^2\pmb{m}^2$ and $\pmb{m}^2$ condense.  
By comparing the numerical results in Figs.~\ref{fig:xy}a and b with Tab.~\ref{tab:order_para}, we conclude the perturbed $\mathbb{Z}_4$ QD model is in the $\mathbb{Z}_4$ QD when $h_w=h_z<h_c$ and trivial phases when $h_w=h_z>h_c$.

Since the Ising* lines meet at the multi-critical point, a natural expectation of the effective theory describing the physics near and at the multi-critical point is the O(2) $\phi^4$ field theory, given by the Lagrangian $\mathcal{L}=\left(\partial \pmb{\phi}\right)^2/2+m^2\pmb{\phi}^2/2+g\pmb{\phi}^4/(4!)+\cdots$, where $\pmb{\phi}=(\phi_0,\phi_1)$ is a two-component order parameter field corresponding to the condensation order parameters $\langle\pmb{e}^2\pmb{m}^2\rangle$ and $\langle\pmb{e}^2\rangle$. 
From the lattice model perspective, the duality transformation gauges the $\mathbb{Z}_2\times\mathbb{Z}_2$ symmetry of the two-coupled Ising model, such that the perturbed $\mathbb{Z}_4$ QD model is obtained. We should also gauge the $\mathbb{Z}_2$ symmetry $\phi_i\leftrightarrow-\phi_i$ in the O(2) $\phi^4$ field to get the effective field theory describing the multi-critical point of the perturbed $\mathbb{Z}_4$ QD model, and the gauging orbifolds the $\phi_i$ fields, i.e., $\phi_i$ can only appear in pairs and $\phi_i$ and $-\phi_i$ become identical~\cite{Universal_CFT_spec_2016}. Therefore, the multi-critical point of the two-coupled Ising (perturbed $\mathbb{Z}_4$ quantum double) model corresponds to the Wilson-Fisher fixed point of the ungauged (gauged) O(2) $\phi^4$ field theory, namely the 3D XY (XY*) conformal field theory~\cite{Wilson_fisher_1972,O_N_CFT_2023}, see SM~\cite{appendix} for more discussions on the distinctions between the XY and XY* theories.
We can predict critical exponents of the condensation order parameters and the area law coefficients from the field theory~\cite{appendix}. 
In Fig.~\ref{fig:xy}c, the area law coefficients match the correlation length critical exponent $\nu_{+}=1/(3-\Delta_{+})$, where $\Delta_{+}$ is the scaling dimension of the relevant field $\pmb{\phi}^2=\phi_0^2+\phi_1^2$ deriving to the gapped phases along $h_w=h_z$.
Fig.~\ref{fig:xy}d shows that the three condensation order parameters exhibit two distinct critical exponents $\beta_{\phi}=\Delta_{\phi}\nu_{+}$ and $\beta_{-}=\Delta_{-}\nu_{+}$, where $\Delta_{\phi}$ and $\Delta_{-}$ are the scaling dimensions of the fields $\phi_{0}$ (or $\phi_{1}$) and $\phi_0^2-\phi_1^2$, which correspond to $\langle\pmb{e}^2\pmb{m}^2\rangle$ (or $\langle\pmb{e}^2\rangle$) and $\langle\pmb{m}^2\rangle$, separately. The numerical results imply that the multi-critical point belongs to the XY* universality class.

\textbf{Continuous topological phase transition between the toric code and double semion phases.}
Finally, we investigate the path $h_x=-1,h_w=2h_{c}-h_z$ [line (ii) in Fig.~\ref{fig:phasediagram}],
which traverses the TC phase, the DS phase, and the multi-critical point, and is orthogonal to $h_w=h_z$ [line (i)]. 
The area law coefficients in Fig.~\ref{fig:xy}e indicate that $\pmb{e}$ and $\pmb{e}^3$ are confined when $h_w>h_z$ and $\pmb{e}^n\pmb{m}^k$ with $n,k\in\{1,3\}$ are confined when $h_w<h_z$, while $\pmb{m}$ and $\pmb{m}^3$ are confined whenever $h_w\neq h_z$. According to the condensation order parameters in Fig.~\ref{fig:xy}f, $\pmb{e}^2\pmb{m}^2$ condenses when $h_w>h_z$, while $\pmb{e}^2$ condenses when $h_w<h_z$. By comparing the numerical results in Figs.~\ref{fig:xy}e and f with Tab.~\ref{tab:order_para}, we conclude that along the line (ii), the system is in the TC phase when $h_z>h_w$ and in the DS phase when $h_w>h_z$.

As shown in Fig.~\ref{fig:xy}g, we find that the area law coefficients exhibit a critical exponent consistent with the correlation length exponent $\nu_{-}=1/(3-\Delta_{-})$, where $\Delta_{-}$ is the scaling dimension of the relevant field $\phi_0^2-\phi_1^2$ deriving the multi-critical point to the TC or DS phases. Fig.~\ref{fig:xy}h shows that the condensation order parameter exhibits the critical exponent $\beta_{\phi}'=\Delta_{\phi}\nu_{-}$. The distinct critical exponents from the lines (i) and (ii) in Fig.~\ref{fig:phasediagram} further confirm that the multi-critical point belongs to the XY* universality class, and a detailed field theory analysis is provided in SM~\cite{appendix}.

Interestingly, the continuous TC-DS transition exhibits features that resemble deconfined quantum critical points~\cite{DQCP_2004,DQCP_PRB_2004}—a direct continuous transition between two phases that spontaneously break different symmetries and
exhibit fractionalized excitations—in the following aspects: First, from Fig.~\ref{fig:xy}f,  $\langle\pmb{e}^2\rangle=0$ in the DS phase while $\langle\pmb{e}^2\pmb{m}^2\rangle=0$ in the TC phase, implying the gapped TC and DS phases spontaneously break two distinct 1-form symmetries, $\prod_{e\in C}Z_e^2$ and $\prod_{e\in \hat{C}}W_e^2$, respectively.  Second, the anyons $\pmb{m}$ and $\pmb{m}^3$ are confined in both the TC and DS phases, but they are deconfined at the continuous TC-DS transition point, indicated by the area law coefficient $\alpha(\pmb{m})$ shown in Fig.~\ref{fig:xy}e; since $\pmb{m}\times\pmb{m}=\pmb{m}^3\times\pmb{m}^3=\pmb{m}^2$ is a well-defined anyon in both TC and DS phases, $\pmb{m}$ and $\pmb{m}^3$ are fractionalized anyons with respect to the $\mathbb{Z}_2$ topological order. A difference between the continuous TC-DS transition point and deconfined quantum critical points is that two broken 1-form symmetries $\prod_{e\in C}Z_e^2$ and $\prod_{e\in \hat{C}}W_e^2$ do not have mixed anomalies.

\textbf{Discussions and outlook.} 
In this letter, we propose a perturbed $\mathbb{Z}_4$ QD model to study the topological phase transitions between the TC and DS phases, from which we find that a direct continuous phase transition exists between these two $\mathbb{Z}_2$ topologically ordered phases.
We identify that the continuous TC-DS transition belongs to the XY* universality class, where deconfined $\mathbb{Z}_4$ anyons emerge. Our findings imply that if we consider a model without any explicit $\mathbb{Z}_4$ structure but exhibiting a continuous TC-DS transition, fractionalized anyons can emerge at the transition.
We emphasize that the TC-DS transition is not a fine-tuned result, since it is a continuous transition line in the plane defined by $h_x<0.8$ and $h_z = h_w$, and we further verified this conclusion through the analysis of O(2) field theory~\cite{appendix}.
Since the TC and DS phases are dual to two different $\mathbb{Z}_2$ symmetry protected topological (SPT) phases in $(2+1)$D~\cite{Levin_Gu_2012}, the multi-critical point can also be interpreted as a topological critical point between two $\mathbb{Z}_2$ SPT phases. We note that a topological phase transition with an exact continuous symmetry can realize a deconfined quantum critical point~\cite{Nat_pivot_DQCP_2023}. It is therefore interesting to consider a model that exhibits the continuous TC–DS transition while maintaining an exact continuous symmetry at the critical point.
Our results shed light on constructing a general theoretical framework for continuous topological phase transitions beyond anyon condensation, i.e. considering emergent anyons to reconcile two incompatible topological orders. 

Since the TC–DS transition can be simulated on a digital quantum computer~\cite{Leo_2024}, it would be interesting to experimentally realize this transition and measure the corresponding anyonic order parameters. In addition, one may consider adding more perturbations like $-h_x' \sum_e X_e + \text{h.c.}$, $-h_z' \sum_e Z_e + \text{h.c.}$
to the perturbed $\mathbb{Z}_4$ QD model in Eq.~\eqref{eq:Z4_QD_H}. 
These terms explicitly break the exact 1-form symmetries $\prod_{e\in C}Z_e^2$, $\prod_{e\in \hat{C}}X_e^2$, hence the duality transformation and non-local order parameters do not work anymore. In such cases, the Fredenhagen–Marcu order parameter can be used to detect anyon condensation~\cite{FM_1983,FM_1986,xu_FM_2024}, and an adiabatic approach can be used to investigate confinement~\cite{xu_roughen_2025}. 
Alternatively, the broken exact 1-form symmetry could be restored via quantum error correction~\cite{QEC_1_form_2025}.

\textbf{Acknowledgments.}
The authors appreciate valuable discussions with L. Wang on numerical computations and support in providing computational resources. W.-T. Xu thanks R.-Z. Huang for helpful comments and acknowledges support from the Munich Quantum Valley, which is supported by the Bavarian
state government with funds from the Hightech Agenda Bayern Plus.

\textbf{Data availability.}
The code and data are available on Zenodo~\cite{zenodo_2025}.
%


\begin{widetext}

\setcounter{table}{0}
\setcounter{figure}{0}
\setcounter{equation}{0}
\setcounter{section}{0}
\renewcommand{\theequation}{S\arabic{equation}}
\renewcommand{\thefigure}{S\arabic{figure}}
\renewcommand{\thetable}{S\arabic{table}}
\renewcommand{\thesection}{S\Roman{section}}

\section{Anyon condensation theory of the $\mathbb{Z}_4$ quantum deouble model}\label{app:Z4_anyon_cond}

In this section, we review the properties of anyons in the $\mathbb{Z}_4$ topological order, and apply the anyon condensation theory to the $\mathbb{Z}_4$ topological order. There are three types of anyons in the $\mathbb{Z}_4$ topological phase. The electric charges \(\{\pmb{e}, \pmb{e}^2, \pmb{e}^3\}\) correspond to eigenvalues \(\{\mathrm{i}, -1, -\mathrm{i}\}\) of \(A_v\), while the magnetic fluxes \(\{\pmb{m}, \pmb{m}^2, \pmb{m}^3\}\) correspond to the same eigenvalues of \(B_p\). Their composites, \(\pmb{e}^n \pmb{m}^k\) (\(n,k\in\{1,2,3\}\)), are known as dyons. As an Abelian topological order, all anyons of the $\mathbb{Z}_4$ topological order are Abelian anyons with the quantum dimension $1$. The topological spins of the charges and the fluxes are $1$. The topological spins of the dyon $\pmb{e}^n\pmb{m}^k$ are determined by the mutual braiding phase factor $\mathrm{i}^{nk}$ between $\pmb{e}^n$ and $\pmb{m}^k$. The mutual braiding of the anyons $\pmb{e}^n \pmb{m}^k$ and $\pmb{e}^q \pmb{m}^r$ ($n,k,q,r\in\{0,1,2,3\}$) gives rise to the phase factor  $\mathrm{i}^{nr-kq}$. The properties of these anyons are summarized in Tab.~\ref{tab:Anyons_of_Z4}.

\begin{table*}[htbp]
\centering
\caption{Anyons of the $\mathbb{Z}_4$ topological order and condensation patterns in the TC and DS phases. The abbreviation ``conf.'' denotes confinement, and ``cond.'' denotes condensation.
}
\begin{tabular}{C{2.5cm} C{0.2cm}
C{0.7cm} C{0.2cm}
C{0.7cm} C{0.7cm} C{0.7cm} C{0.2cm}
C{0.7cm} C{0.7cm} C{0.7cm} C{0.2cm}
C{0.7cm} C{0.7cm} C{0.7cm} C{0.7cm} C{0.7cm} C{0.7cm} C{0.7cm} C{0.7cm} C{0.7cm}
}
\toprule
Type && 
Trivial &&  
\multicolumn{3}{c}{Charge} & & 
\multicolumn{3}{c}{Flux} & & 
\multicolumn{9}{c}{Dyon}  \\
Anyon & & 
$\pmb{1}$ & & 
$\pmb{e}$ & $\pmb{e}^2$ & $\pmb{e}^3$ & &
$\pmb{m}$ & $\pmb{m}^2$ & $\pmb{m}^3$ & & 
$\pmb{e}\pmb{m}$ &$\pmb{e}\pmb{m}^2$ &$\pmb{e}\pmb{m}^3$ & $\pmb{e}^2\pmb{m}$ & $\pmb{e}^2\pmb{m}^2$ & $\pmb{e}^2\pmb{m}^3$ &$\pmb{e}^3\pmb{m}$ &$\pmb{e}^3\pmb{m}^2$ & $\pmb{e}^3\pmb{m}^3$\\

Topological spin  & & 
1  && 
1 & 1 & 1 & & 
1 & 1 & 1 & & 
i & -1 & -i & -1 & 1 & -1 & -i & -1 & i \\
\midrule
TC && 
$\pmb{1}$ && 
$\pmb{e}$ & cond. & $\pmb{e}$ && 
conf. & $\pmb{m}^2$ & conf. && 
conf. & $\pmb{e}\pmb{m}^2$ & conf. & conf. & $\pmb{m}^2$ & conf. & conf. & $\pmb{e}\pmb{m}^2$ & conf\\
DS && 
$\pmb{1}$ && 
conf. & $\pmb{m}^2$ & conf. && 
conf. & $\pmb{m}^2$ & conf. && 
$\pmb{e}\pmb{m}$ & conf. & $\pmb{e}\pmb{m}^3$  & conf. & cond. & conf. & $\pmb{e}\pmb{m}^3$ & conf.  & $\pmb{e}\pmb{m}$ \\
\bottomrule
\end{tabular}
\label{tab:Anyons_of_Z4}
\end{table*}

Then we consider anyon condensation in the $\mathbb{Z}_4$ topological order. The basic rules for Abelian anyon condensation are ~\cite{Anyon_cond_theory_2002_app, Anyon_cond_theory_2009_app, Liang_kong_2014_app, anyon_cond_2018_app}: 
\begin{itemize}
\setlength{\itemsep}{0.0em}
\setlength{\parskip}{0.1em}
\item Only bosonic anyons whose topological spins are $1$ can condense.
\item Anyons whose mutual braiding phase factors with the condensed anyons are not $1$ are confined.
\item Two anyons that differ by a condensed anyon via the fusion rules become identical.
\end{itemize}
Based on these rules, we can find that by condensing $\pmb{e}^2$, the remaining anyons are $\{\pmb{1},\pmb{e},\pmb{m}^2,\pmb{e}\pmb{m}^2\}$, which are exactly the anyons of the toric code, as shown in Tab.~\ref{tab:Anyons_of_Z4}. A different toric code can be obtained by condensing $\pmb{m}^2$ such that the remaining anyons are $\{\pmb{1},\pmb{m},\pmb{e}^2,\pmb{e}^2\pmb{m}\}$. Moreover, there is a bosonic dyon $\pmb{e}^2\pmb{m}^2$ in the $\mathbb{Z}_4$ topological order. By condensing $\pmb{e}^2\pmb{m}^2$, the remaining anyons are $\{\pmb{1},\pmb{e}\pmb{m},\pmb{e}\pmb{m}^3,\pmb{e}^2\}$, also in Tab.~\ref{tab:Anyons_of_Z4}. These anyons are exactly the anyons $\{\pmb{1},\pmb{s},\bar{\pmb{s}},\pmb{b}=\pmb{s}\bar{\pmb{s}}\}$  of the DS, where $\pmb{s}$ denotes the semion.

\section{Derivation of the duality transformation}\label{app:duality}

\subsection{Duality transformation between Hamiltonians}
In this section, we present a detailed derivation of the duality transformation and discuss the underlying physics. 
The first step is to map the $\mathbb{Z}_4$ qudit operators on the edges of the lattice to the new $\mathbb{Z}_4$ qudit operators on the vertices and plaquettes
(see Fig.~2b in the main text):
\begin{equation}\label{eq:duality_step_1}
     A_v\rightarrow X_v,\,\, B_p\rightarrow Z_p,\,\, X^2_e\rightarrow X^2_{p(e)}X^2_{p'(e)},\,\,Z^2_e\rightarrow Z^2_{v(e)}Z^2_{v'(e)},
 \end{equation}
 where $p(e)$ and $p'(e)$ [$v(e)$ and $v'(e)$] denote a pair of plaquettes [vertices] that are nearest neighboring to the edge $e$. This step takes advantage of conserved quantities of $H_{\text{QD}}$, and it can be checked that the set of operators $\{A_v,B_p,X_e,Z_e|A^2_v=B_p^2=1\}$ and $\{X_v,Z_p,X_pX_{p'},Z_vZ_{v'}|X_v^2=Z_p^2=1\}$ satisfy the same algebra.
Applying the transformation in Eq.~\eqref{eq:duality_step_1} to the perturbed $\mathbb{Z}_4$ QD model, we get:
\begin{align}\label{eq:Z4_spin_H}
    H_{\text{QD}}&=-\left[\sum_v X_v+\sum_p Z_p+\frac{1}{2}\sum_{p}\left(X_{v(p)}Z_p+X_{v(p)}Z^{\dagger}_p\right)+\text{h.c.}\right]-h_x\sum_{\langle pp'\rangle} X^2_pX_{p'}^2-h_z\sum_{\langle vv'\rangle} Z^2_vZ_{v'}^2 -h_w\sum_{\langle pp'\rangle} X^2_pX^2_{p'}Z^2_{v(p)}Z^2_{v(p')}, 
\end{align}
where $v(p)$ denotes the vertex $v$ adjacent to the plaquette $p$ and in the southwest direction of $p$, as shown in Fig.~\ref{fig:anyon_order_para}c.

The above model can be further simplified by expressing the $\mathbb{Z}_4$ operators in terms of two sets of $\mathbb{Z}_2$ Pauli matrices $\{\sigma^x,\sigma^z\}$ and $\{\tau^x,\tau^z\}$:
\begin{align}
    X_v+X_v^\dagger=(\mathbbm{1}_v+\sigma_v^x)\otimes \tau_v^x,\quad  Z_p+Z_p^\dagger=\sigma_p^z\otimes(\mathbbm{1}_p+\tau_p^z),\quad
    Z_v^2=\mathbbm{1}_v\otimes \tau_v^z,\quad
    X_p^2= \sigma_p^x\otimes\mathbbm{1}_p,\quad 
    X_v^2= \sigma_v^x\otimes \mathbbm{1}_v,\quad
    Z_p^2= \mathbbm{1}_p\otimes \tau_p^z.
\end{align}
 
Due to the constraint $X_v^2=Z_p^2=1$, we have $\sigma_v^x=\tau_p^z=1$, which further simplifies the relations to:
\begin{align}\label{eq:Z2_spin_H}
      X_v+X_v^\dagger&=2\tau_v^x,\quad  
      Z_p+Z_p^\dagger=2\sigma_p^z,\quad  
      Z_v^2=\tau_v^z,\quad 
      X_p^2= \sigma_p^x.
\end{align}

Then, the Hamiltonian in Eq.~\eqref{eq:Z4_spin_H} can be rewritten in terms of these Pauli matrices:
\begin{align}
    H_{\text{dual}}&=-2\left(\sum_v \tau^x_v+\sum_p \sigma_p^z+\sum_{p}\tau^x_{v(p)}\sigma^z_p\right)-h_x\sum_{\langle pp'\rangle} \sigma^x_p\sigma^x_{p'}-h_z\sum_{\langle vv'\rangle} \tau^z_v\tau_{v'}^z-h_w\sum_{\langle pp'\rangle} \sigma^x_p\sigma^x_{p'}\tau^z_{v(p)}\tau_{v(p')}^z.
\end{align}
By shifting the dual lattice such that it overlaps with the primal lattice, and relabeling $v,v'$ and $p,p'$ using $i,j$, we obtain the two-coupled $(2+1)$D quantum Ising model: 
\begin{align}\label{eq:supp_at}
    H_{\text{dual}}=-2\sum_i \left(\tau^x_i+ \sigma_i^z+\tau^x_i\sigma^z_i\right)-h_x\sum_{\langle ij\rangle} \sigma^x_i\sigma^x_{j}-h_z\sum_{\langle ij\rangle} \tau^z_i\tau_{j}^z-h_w\sum_{\langle ij\rangle} \tau^z_i\tau_{j}^z\sigma^x_i\sigma^x_{j}.
\end{align}
We validate the duality transformation by comparing the energy spectra of the perturbed $\mathbb{Z}_4$ QD model $H_{\text{QD}}$ and the dual model $H_{\text{dual}}$ using exact diagonalization. The agreement between the spectra of the two models confirms the correctness of the duality mapping.

In certain special limits, the two-coupled quantum Ising model in Eq.~(\ref{eq:supp_at}) can be further simplified.
When $h_x=h_w=0$, the degrees of freedom of $\{\sigma_i^z\}$ are fixed to $1$. The model reduces to a single $(2+1)$D transverse-field Ising model (TFIM) : $H=-4 \sum_i \tau_i^x - h_z \sum_{\langle ij\rangle} \tau_i^z \tau_j^z$. This yields the critical point at $h_z = 4 / h_c^{\text{TFIM}}$, where $ h_c^{\text{TFIM}} = 3.044\,330(6)$ is the known critical field of the TFIM~\cite{youjin_2020_app}.
When $h_x=0,~h_w\rightarrow +\infty$, there are constraints $\tau_i^z \tau_j^z=\sigma_i^x \sigma_j^x=1, \forall \langle i,j\rangle$. $H_{\text{dual}}$ reduces to another TFIM: $H=-2\sum_i s_i^x - h_z \sum_{\langle ij\rangle} s^z_i s^z_j$, where $s_i^z=\tau_i^z$ and $s_i^x=\tau_i^x \sigma_i^z$.
Similarly, when $h_x=h_z=0$ and $h_x=0,~h_z\rightarrow +\infty$, $H_{\text{dual}}$ also reduces to the TFIM.

Our duality transformation at these special limits are almost the Wegner duality transformation, which is a non-invertible transformation, and it does not preserve the locality of operators that are not 1-form symmetric. The Wegner duality exactly transforms the $\mathbb{Z}_2$ toric code model in a transverse/longitudinal field on a torus to the direct sum of the even sector TFIM with the periodic boundary condition and the even sector TFIM with the anti-periodic boundary condition~\cite{Universal_CFT_spec_2016_app}. To show our duality transformation does not change the critical behavior of the order parameters we studied in the main text, we check how our duality transforms the symmetry sectors and the boundary conditions between the perturbed $\mathbb{Z}_4$ QD model and the two-coupled Ising model.

Let us first consider the symmetry sectors. From Eqs.~\eqref{eq:duality_step_1} and \eqref{eq:Z2_spin_H}, we know $\sigma_i^z$ and $\tau_i^x$ of the two-coupled Ising model correspond to $B_p^2$ and $A_v^2$ of the perturbed $\mathbb{Z}_4$ QD model. When putting the system on a torus, the $\mathbb{Z}_2\times \mathbb{Z}_2$ symmetry operators of the two-coupled Ising model satisfy $\prod_i\sigma_i^z=\prod_pB^2_p=\prod_i\tau_i^x=\prod_vA^2_v=1$, therefore, the perturbed $\mathbb{Z}_4$ QD model maps only to the even sectors of the two $\mathbb{Z}_2$ symmetries.

Next, let us consider the boundary condition.  In the perturbed $\mathbb{Z}_4$ QD model, besides the conserved quantities $A_v^2$ and $B_p^2$, the following non-contractible loop operators are also independent conserved quantities for the system on a torus:  
\begin{equation}
\prod_{e\in \hat{C}_x}X_e^2,\quad \prod_{e\in \hat{C}_y}X_e^2,\quad \prod_{e\in C_x}Z_e^2,\quad \prod_{e\in C_y}Z_e^2,
\end{equation}
where $\hat{C}_{x(y)}$ are non-contractible loops on the dual lattice in $x(y)$ direction, and  $C_{x(y)}$ are non-contractible loops on the primal lattice in $x(y)$ direction. Again using  Eqs.~\eqref{eq:duality_step_1} and \eqref{eq:Z2_spin_H} we find $\prod_{e\in \hat{C}_x}X_e^2=\prod_{e\in \hat{C}_y}X_e^2=\prod_{e\in C_x}Z_e^2=\prod_{e\in C_y}Z_e^2=1$, so the two-coupled Ising model with the periodic boundary condition only corresponds to the perturbed $\mathbb{Z}_4$ QD model in the even sector of these non-contractible 1-form symmetry operators. Actually, some eigenstates (including ground states) of the perturbed $\mathbb{Z}_4$ QD model are eigenstates of some non-contractible 1-form symmetry operators with the eigenvalue $-1$. Therefore, to obtain the odd sectors of these non-contractible 1-form symmetry operators, we can slightly modify the duality transformation in Eq.~\eqref{eq:duality_step_1}:
\begin{equation}
   X_e^2 =
\begin{cases}
X^2_{p(e)}X^2_{p'(e)}, & \text{if } e\notin \hat{C}_x\,\, \text{and}\,\, \hat{C}_y,\\
\pm X^2_{p(e)}X^2_{p'(e)}, & \text{if } e\notin \hat{C}_x\,\, \text{or}\,\, \hat{C}_y,
\end{cases}\quad\quad
Z_e^2 =
\begin{cases}
Z^2_{v(e)}Z^2_{v'(e)}, & \text{if } e\notin C_x\,\, \text{and}\,\, C_y,\\
\pm Z^2_{v(e)}Z^2_{v'(e)}, & \text{if } e\notin C_x\,\, \text{or}\,\, C_y.
\end{cases}
\end{equation}
In the case with the $+(-)$ sign before $X^2_{p(e)}X^2_{p'(e)}$ or $Z^2_{v(e)}Z^2_{v'(e)}$ in the second line, the perturbed $\mathbb{Z}_4$ QD model is mapped to the the two-coupled Ising model with the periodic (anti-periodic) boundary condition for the $\sigma^x$ or $\tau^z$ spins. Here, the anti-periodic boundary condition means that the ferromagnetic couplings along some non-contractible loops are changed to the anti-ferromagnetic couplings. In summary, the complete duality transformation of the perturbed $\mathbb{Z}_4$ QD model $H_{\text{QD}}$ on a torus is
\begin{equation}\label{eq:complete_duality}
H_{\text{QD}}=\bigoplus_{b^{\sigma}_x,b^{\sigma}_y,b^{\tau}_x,b^{\tau}_y} \frac{1+\prod_i \sigma^z_i}{2}\frac{1+\prod_i \tau^x_i}{2}H_{\text{dual}}(b^{\sigma}_x,b^{\sigma}_y,b^{\tau}_x,b^{\tau}_y),
\end{equation}
where $b^{\sigma}_x,b^{\sigma}_y,b^{\tau}_x,b^{\tau}_y$ are boundary conditions in $x$ and $y$ directions of the $\sigma$ or $\tau$ spins which can be either periodic ($+$) or antiperiodic ($-$). The ground state of $\frac{1+\prod_i \tau^x_i}{2}\frac{1+\prod_i \sigma^z_i}{2}H_{\text{dual}}(+,+,+,+)$ is always non-degenerate, and its energy is always equal or lower than ground state energies of all the other fifteen Hamiltonians in the direct sum. So the ground state of $\frac{1+\prod_i \tau^x_i}{2}\frac{1+\prod_i \sigma^z_i}{2}H_{\text{dual}}(+,+,+,+)$ corresponds to one of the ground state of the perturbed $\mathbb{Z}_4$ QD model which satisfy $\prod_{e\in \hat{C}_x}X_e^2=\prod_{e\in \hat{C}_y}X_e^2=\prod_{e\in C_x}Z_e^2=\prod_{e\in C_y}Z_e^2=1$. In the next subsection, we will show that it is valid to evaluate the order parameters of the $H_{\text{QD}}$ using the ground state $H_{\text{dual}}$ regardless of the even sector projectors before it.

\begin{figure*}[htbp]
    \centering
    \includegraphics[width=0.70\linewidth]{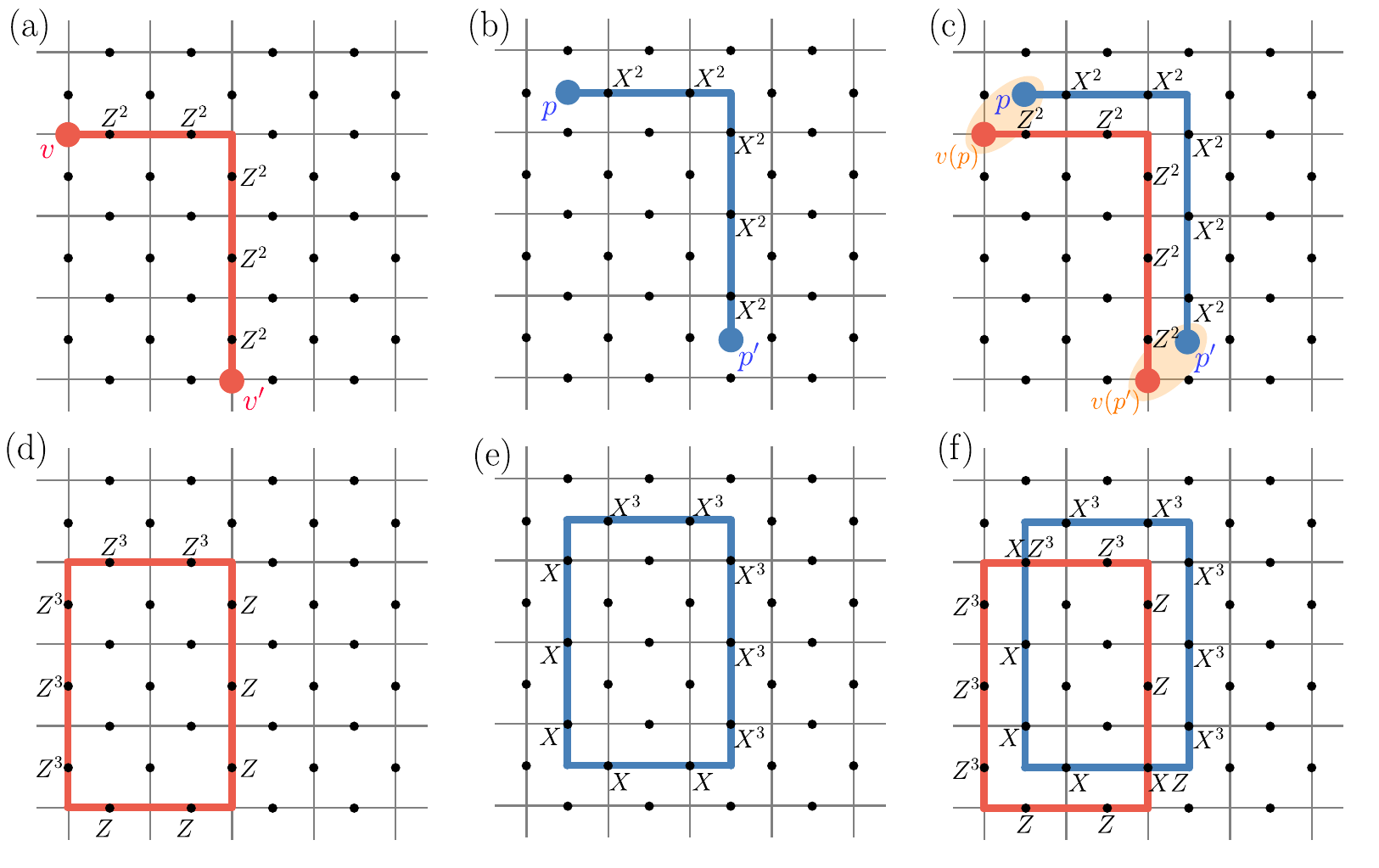}
    \caption{\textbf{String and loop operators defining the condensation order parameters and the Wilson loop order parameters}. (a) Sting operator creating two $\pmb{e}^2$ at $v$ and $v'$. (b) Sting operator creating $\pmb{m}^2$ at $p$ and $p'$. (c) Sting operator creating $\pmb{e}^2\pmb{m}^2$ at $p$ and $p'$. (d) Wilson loop operator associated to $\pmb{e}$. (e) Wilson loop operator associated to $\pmb{m}$. (f) Wilson loop operator associated to $\pmb{e}\pmb{m}$.}
    \label{fig:anyon_order_para}
\end{figure*}

\subsection{Duality transformation between order parameters}

Using the duality transformation, the condensation order parameter and Wilson loop order parameters associated with the 1-form symmetries of the $\mathbb{Z}_4$ QD model can be mapped to the order and disorder parameters of the two-coupled Ising model with the $\mathbb{Z}_2\times \mathbb{Z}_2$ 0-form symmetries, respectively. Specifically, using the relations $X^2_e\rightarrow \sigma^x_{p(e)}\sigma^x_{p'(e)},\,\,Z^2_e\rightarrow \tau^z_{v(e)}\tau^z_{v'(e)}$, where $p(e)$ [$v(e)$] denotes the plaquette [vertex] adjacent to the edge $e$, one can easily check that the string operators transform as:
\begin{equation}
    \prod_{e\in \hat{S}(p,p')}X_e^2\rightarrow \sigma^x_p\sigma^x_{p'},\quad \prod_{e\in S(v,v')}Z_e^2\rightarrow \tau^z_v\tau^z_{v'},\quad\prod_{e \in \hat{S}(p, p')} W_e^2 \longrightarrow \sigma^x_p \sigma^x_{p'} \tau^z_{v(p)} \tau^z_{v(p')},
\end{equation}
where $S(v,v')$ [$\hat{S}(p,p')$] denotes the string on the primal [dual] lattice whose two end points are at two vertices [plaquettes] $v$ [$p$] and $v'$ [$p'$], 
see Figs.~\ref{fig:anyon_order_para}a-c. Furthermore, using the relations  $A_v+A_v^{\dagger}\rightarrow 2\tau_i^x,\,\, B_p+B_p^{\dagger}\rightarrow 2\sigma_i^z$, we can transform the Wilson loop operators of the $\mathbb{Z}_4$ QD model to the disorder operators of the dual model~\cite{Kadanoff_1971_app,Fradkin2017_app}: 
\begin{align}\label{eq:Wilson_loop}
    &\frac{1}{2}\left[\left(\prod_{e\in C}Z^{\gamma(e)}_e\right)+\text{h.c.}\right] 
    &&= \frac{1}{2}\left[\left(\prod_{p\in a_C}B_p\right)+\text{h.c.}\right] 
    &&\xrightarrow{\makebox[1.3cm]{\smash{\scriptsize $B_p^2 = 1$}}}&&
    \frac{1}{2^{|a_C|}} \prod_{p\in a_C} \left(B_p + B^{\dagger}_p\right) 
    &&\longrightarrow&& \prod_{i\in R}\sigma^z_i, \notag \\
    &\frac{1}{2}\left[\left(\prod_{e\in \hat{C}}X^{\gamma(e)}_e\right)+\text{h.c.}\right] 
    &&= \frac{1}{2}\left[\left(\prod_{v\in a_{\hat{C}}}A_v\right)+\text{h.c.}\right] 
    &&\xrightarrow{\makebox[1.3cm]{\smash{\scriptsize $A_v^2 = 1$}}}&&
    \frac{1}{2^{|a_{\hat{C}}|}} \prod_{v\in a_{\hat{C}}} \left(A_v + A^{\dagger}_v\right) 
    &&\longrightarrow&& \prod_{i\in R}\tau^x_i, \notag \\
    &\frac{1}{2}\left[\left(\prod_{e\in \hat{C}}X^{\hat{\gamma}(e)}_e Z^{\gamma(e')}_{e'}\right)+\text{h.c.}\right] 
    &&=\frac{1}{2}\left[\left(\prod_{p\in a_{C}}A_{v(p)}B_p\right)+\text{h.c.}\right] 
    &&\xrightarrow{\makebox[1.3cm]{\smash{\scriptsize $A_v^2 = B_p^2 = 1$}}}&&
    \frac{1}{4^{|a_C|}} \prod_{p\in a_{C}} \left(A_{v(p)}B_p + A^{\dagger}_{v(p)}B_p + \text{h.c.}\right) 
    &&\longrightarrow&& \prod_{i\in R} \tau^x_i\sigma^z_i,
\end{align}
where $\gamma(e),~\hat{\gamma}(e)$ taking values $1$ or $3$ depends on the direction of the edge $e$, see Figs.~\ref{fig:anyon_order_para}d-f.
The symbol $a_C$ ($a_{\hat{C}}$) is the set of plaquettes (vertices) surrounded by $C$ ($\hat{C}$). $R$ denotes the set of sites surround by the loop $C$ ($\hat{C}$) after shifting the dual lattice to the primal lattice.   
The derivation takes advantage of the fact that the ground states of the perturbed $\mathbb{Z}_4$ QD model satisfy $A_v^2=B_p^2=1$.

There is a small subtly when evaluating the duality transformed order parameters. When the two coupled Ising model $H_{\text{dual}}$ is in a symmetry-breaking phase, the even sector projectors enforce the ground states to be the symmetric cat states. Because of the relation $\lim_{|p-p'|\rightarrow \infty}|\langle\prod_{e\in\hat{S}(p,p')}X^2_e\rangle|^{1/2}=\lim_{|i-j|\rightarrow \infty}|\langle \sigma^x_i\sigma^x_j\rangle_{\text{cat}}|^{1/2}=\lim_{|i-j|\rightarrow \infty}|\langle \sigma^x_i\sigma^x_j\rangle|^{1/2}=|\langle \sigma^x_i\rangle|$ (similar for $\tau^z$), where the string operator is evaluated using a ground state of the perturbed $\mathbb{Z}_4$ QD model and the $\mathbb{Z}_2$ Pauli operators are evaluated using a ground state of the two-coupled Ising model (``cat'' denotes the symmetrized ground state in the even sectors of two $\mathbb{Z}_2$ symmetries), it is still valid to evaluate the string order parameter of the perturbed $\mathbb{Z}_4$ QD model using the symmetry breaking iPEPS of the two-coupled Ising model. Moreover, from Eq.~\eqref{eq:Wilson_loop}, the Wilson loop operators of the perturbed $\mathbb{Z}_4$ quantum double model are mapped to the disorder operators of the two-coupled Ising model, and it can be checked that $\langle\prod_{i\in R}\sigma^z_i\rangle_{\text{cat}}=\langle\prod_{i\in R}\sigma^z_i\rangle$ (similar for $\tau^x$), so it does not matter if we use the symmetric cat states or symmetry breaking states.

Since the duality transformation maps between two sets of order parameters related to the spontaneous 1-form symmetry breaking and the spontaneous 0-form symmetry breaking, separately, we can know the mapping between the 1-form symmetry breaking patterns and the 0-form symmetry breaking patterns. Specifically, the $\mathbb{Z}_4$ QD model has a $\mathbb{Z}^{(1)}_2\times \mathbb{Z}^{\prime(1)}_2$ 1-form symmetry (the superscripts denote the form of the symmetry), which has different spontaneous symmetry breaking patterns in different topologically ordered phases, as shown in Tab.~\ref{tab:SSB_pattern}. 
After the duality transformation, the two-coupled quantum Ising model has a $\mathbb{Z}^{(0)}_2\times \mathbb{Z}^{\prime(0)}_2$ 0-form symmetry, and the spontaneous symmetry breaking patterns in different phases are also summarized in Tab.~\ref{tab:SSB_pattern}.

\subsection{The underlying physics of the duality transformation}
An interesting question regarding the duality transformation is: How can we explicitly construct the operator that implements the duality transformation?
We can think about this question from the fact that two categories $\text{Vec}(\mathbb{Z}_4)$ and $\text{Vec}^{\omega_{\text{II}}}(\mathbb{Z}_2\times\mathbb{Z}_2)$ are Morita equivalent~\cite{PHD_thesis_1995_app,Dominic_2017_app}, where $\omega_{\text{II}}$ is a type-II cocycle of the $\mathbb{Z}_2\times \mathbb{Z}_2$ group. 
Specifically, we can apply a finite depth quantum circuit (FDQC) to transform the perturbed $\mathbb{Z}_4$ QD model to a perturbed type-II cocycle twisted $\mathbb{Z}_2\times \mathbb{Z}_2$ QD model~\cite{FDQC_Morita_2022_app}.
Then, applying the tensor product of two Wegner duality (duality$ ^{\otimes 2}$) transformations on the perturbed but untwisted $\mathbb{Z}_2\times \mathbb{Z}_2$ quantum double model, we get a perturbed model for a non-trivial  $\mathbb{Z}_2\times \mathbb{Z}_2$ symmetry-protected topological (SPT) phase. 
Finally, via another FDQC, we can transform the perturbed non-trivial SPT model to the two-coupled Ising model.  The effect of applying these transformations on the fixed point $\mathbb{Z}_4$ QD model can be summarized as:
\begin{equation}~\label{eq:summary_transformation}
    \mbox{Vec}(\mathbb{Z}_4)\xrightarrow{\text{FDQC}}\mbox{Vec}^{\omega_{II}}(\mathbb{Z}_2\times\mathbb{Z}_2)\xrightarrow{\text{duality} ^{\otimes 2}} \mathrm{nontrivial}\,\, \mathbb{Z}_2\times\mathbb{Z}_2\,\, \mathrm{ SPT} \xrightarrow{\text{FDQC}}\mbox{
    Paramagnet},
\end{equation}
and the transformation for the perturbed $\mathbb{Z}_4$ QD model is the same one.  
 We notice that the DS phase suffers from the intrinsic sign problem~\cite{Hastings_Negative_wavefunction_2015_app,Intrinsic_sign_2020_app,Qi_Zhang_2020_app}, whereas the duality transformation maps it to a paramagnetic phase without the intrinsic sign problem. This indicates that the duality transformation we propose is distinct from the well-known Wegner duality between the $\mathbb{Z}_2$ lattice gauge theory and the Ising model~\cite{Wegner_duality_1971}, which relates two models both without the intrinsic sign problem. Our duality can instead be understood as a combination of the Wegner duality and finite-depth quantum circuits, which together remove the intrinsic sign problem. It would be interesting to explicitly work out the transformations shown in Eq.~\eqref{eq:summary_transformation} in the future.

\begin{table*}[htbp]
\centering
\caption{Symmetry breaking patterns in different topological phases of the perturbed $\mathbb{Z}_4$ QD model, and the corresponding phases and symmetries in the dual two-coupled quantum Ising model. The superscripts denote the form of symmetries.}
\begin{tabular}{C{5.0cm}C{3.0cm}C{3.0cm}C{3.0cm}C{3.0cm}}
\toprule
Perturbed  $\mathbb{Z}_4$ QD model &     $\mathbb{Z}_4$ QD & TC & DS & Trivial \\  

Ground state symmetry& $\mathbb{Z}_1^{(1)}$ & $\mathbb{Z}_2^{(1)}=\{\mathbbm{1},\prod_{e\in C}Z_e^2\}$ & $\mathbb{Z}_2^{\prime(1)}=\{\mathbbm{1},\prod_{e\in \hat{C}}X_e^2 Z_{e'}^2\}$ & $\mathbb{Z}_2^{(1)}\times \mathbb{Z}_2^{\prime(1)}$\\ 
\midrule
Two-coupled quantum Ising model & Paramagnet & $\tau^z$ Ferromagnet & $\sigma^x\tau^z$ Ferromagnet  & Baxter \\  

Ground state symmetry& $\mathbb{Z}_2^{(0)}\times \mathbb{Z}_2^{\prime(0)}$ & $\mathbb{Z}_2^{(0)}=\{\mathbbm{1},\prod_{i}\sigma_i^z\}$ & $\mathbb{Z}_2^{\prime(0)}=\{\mathbbm{1},\prod_{i}\tau_i^x\sigma_i^z\}$ & $\mathbb{Z}_1^{(0)}$\\ 
\bottomrule
\end{tabular}
\label{tab:SSB_pattern}
\end{table*}

\section{Supplemental numerical Results}\label{app:supp_data}

In this work, the ground states of the two-coupled $(2+1)$D quantum Ising model are approximated by $C_{4v}$-symmetric iPEPS~\cite{iPEPS_corboz_2016_app,iPEPS_Laurens_2016_app} with various bond dimensions $D$, using the automatic differentiation technique~\cite{AD_2019_app}. 
The contraction of the squared norm of iPEPS is performed approximately using the $C_{4v}$ symmetric corner transfer matrix renormalization group (CTMRG) algorithm~\cite{CTMRG_1_app,CTMRG_2_app,Fast_CTMRG_app}, with environment bond dimensions $\chi$. 
After evaluating the order parameters and the disorder parameters of the two-coupled quantum Ising model, we map them back to the condensation order parameters and the area law coefficients of the perturbed $\mathbb{Z}_4$ QD model.

\begin{figure}[htbp]
    \centering
    \includegraphics[width=0.82\linewidth]{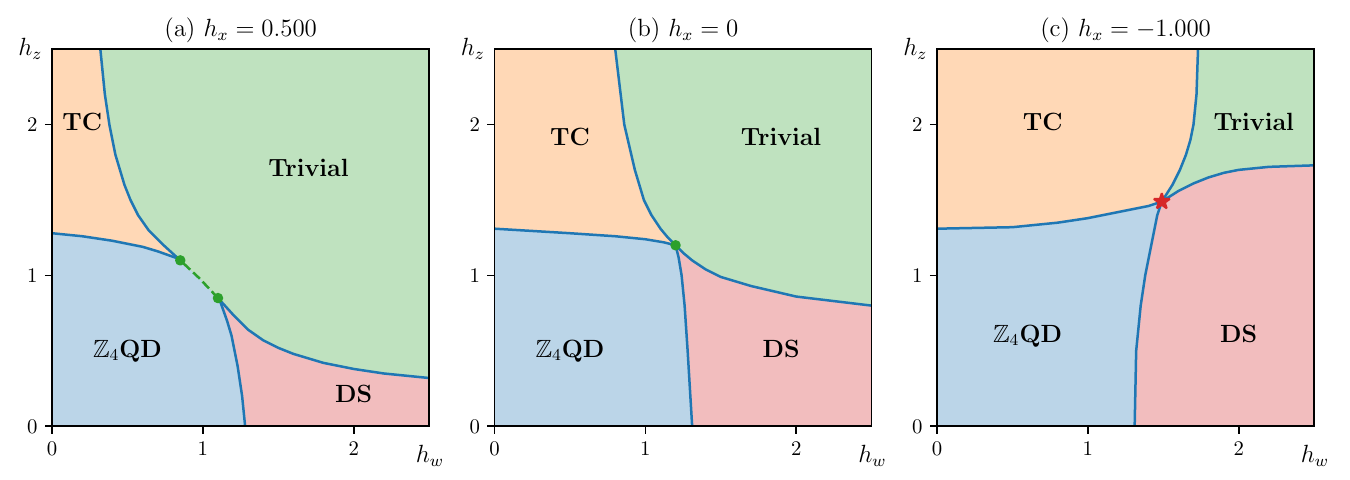}
    \caption{
\textbf{Phase diagrams of the perturbed $\mathbb{Z}_4$ QD model.}
The solid blue lines indicate continuous transitions which belong to the 3D Ising universality class.
The dashed green line and dots represent first-order phase transitions.
The red star marks the multi-critical point that to the XY* universality class.
On the (a) $h_x=0.5$, (b) $h_x=0$, (c) $h_x=-1.0$ plane. 
}
    \label{fig:phasediagrams_supp}
\end{figure}

We show the phase diagrams of the perturbed $\mathbb{Z}_4$ QD model on various $h_x$ planes in Fig.~\ref{fig:phasediagrams_supp}, which exhibit three distinct types of structures.
The phase diagrams for $0.1<h_x<1.0$ resemble Fig.~\ref{fig:phasediagrams_supp}a, where the TC and DS phases are not directly connected and the $\mathbb{Z}_4$ QD and trivial phases are separated by a first-order transition line.
As $h_x$ decreases into the range $-0.2<h_x<0.1$, the first-order line gradually shrinks into a single first-order transition point, such that the four gapped phases meet together, as depicted in Fig.~\ref{fig:phasediagrams_supp}b.
For $-0.8<h_x<-0.2$, numerical results suggest the meeting point of the four phases is associated with a continuous phase transition, but we find that its critical exponents may vary continuously with decreasing $h_x$. Previous studies of the 3D classical AT model suggest that there are several possibilities: weakly first-order~\cite{3D_classical_AT_2012_app}, a wide crossover~\cite{3D_classical_AT_2019_app}, or continuous but non-universal~\cite{3D_classical_AT_2003_app, 3D_classical_AT_2006_app}. So the meeting point in the range $-0.8<h_x<-0.2$ needs further investigation.  
Moreover, in the range $-1.0\leq h_x<-0.8$, the multi-critical point falls within the XY* universality class, and the shape of the phase diagram is simialr to that of $h_x=-1$ plane, as shown in Fig.~\ref{fig:phasediagrams_supp}c. Finally, when $h_x < -1$, corresponding to a strong antiferromagnetic coupling in the two-coupled Ising model, the perturbed $\mathbb{Z}_4$ QD model may exhibit a rich phase diagram, as a mixed-order phase is known to emerge in the 3D classical AT model~\cite{3D_classical_AT_1980_app,3D_classical_AT_2021_app}, which could imply the existence of a phase in the perturbed $\mathbb{Z}_4$ QD model where the $\mathbb{Z}_2$ topological orders of TC and DS coexist.

\begin{figure}[h]
    \centering
    \includegraphics[width=0.62\linewidth]{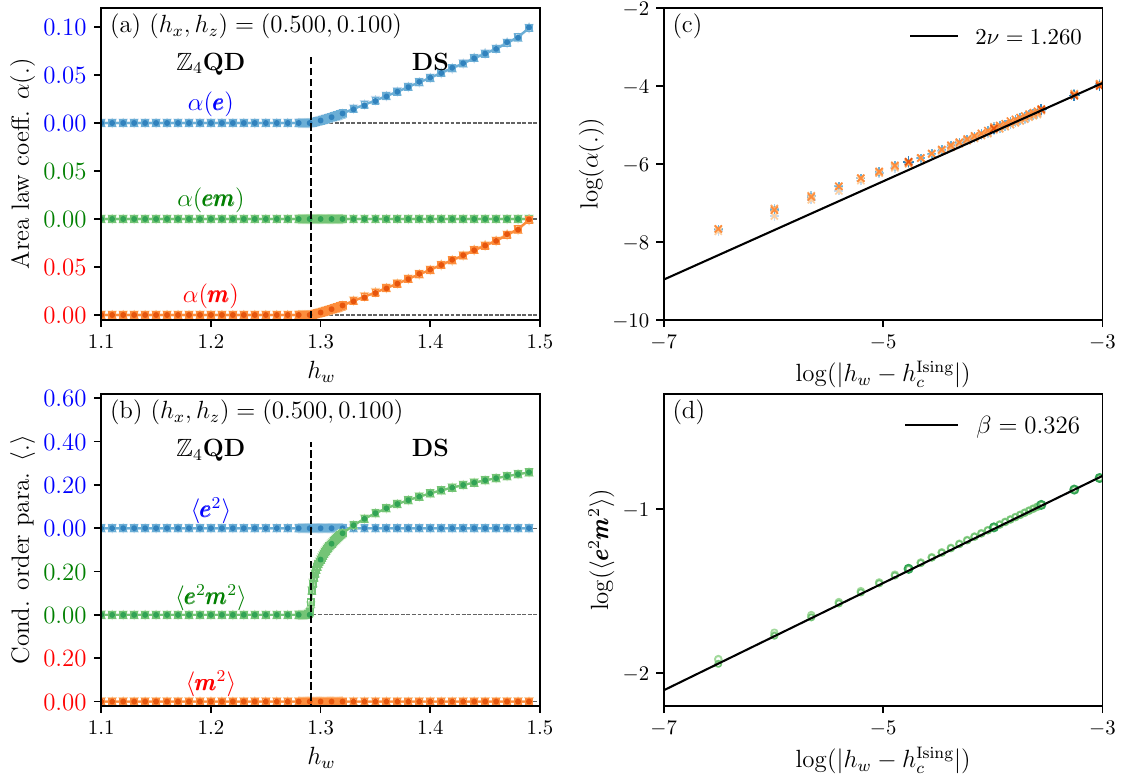}
    \caption{\textbf{Condensation order parameters and area law coefficients along the line $(h_x,h_z)=(0.5,0.1)$ as a function of $h_w$.  } Different colors correspond to different order parameters, and the colors from light to dark correspond to the bond dimensions $(D,\chi)=(5,100),(5,160),(6,100),(7,100)$.
    The vertical black dashed line corresponds to the location of the critical point $h_w \approx 1.292$, denoted by $h_{c}^{\text{Ising}}$.
    The 3D Ising critical exponents for order parameter and correlation length are approximately $\beta\approx 0.326$ and $\nu\approx 0.630$, respectively.
    (a) Area law coefficients.
    (b) Condensation order parameters.
    (c) Double-log plot of the area law coefficients $\alpha(\pmb{e})$ and $\alpha(\pmb{m})$, where the slope of the straight line is $2\nu$.
    (d) Double-log plot of the condensation order parameter $\langle \pmb{e}^2\pmb{m}^2 \rangle$, where the slop of the straight line is $\beta$.
    }
    \label{fig:ising}
\end{figure}

We consider the line $(h_x,h_z)=(0.5,0.1)$ with varying $h_w$, which crosses the $\mathbb{Z}_4$ QD and DS phases in the perturbed $\mathbb{Z}_4$ QD model. 
As shown in Fig.~\ref{fig:ising}, we find that all the physical quantities converge well for iPEPS bond dimension $D = 5$, and further increasing the CTMRG environment bond dimension $\chi$ has a negligible effect.
Figure~\ref{fig:ising}a shows that when $h_w>h_{c}^{\text{Ising}}$, the area law coefficients $\alpha(\pmb{e})>0$ and $\alpha(\pmb{m})>0$ but $\alpha(\pmb{e}\pmb{m})=0$, implying that the charges $\pmb{e}$ and $\pmb{e}^3$ and fluxes $\pmb{m}$ and $\pmb{m}^3$ are confined while the semionic dyon $\pmb{e}^n\pmb{m}^k$ with $n,k\in\{1,3\}$ remains deconfined. 
In Fig.~\ref{fig:ising}b, 
the bosonic dyon $\pmb{e}^2\pmb{m}^2$ condenses ($\langle\pmb{e}^2\pmb{m}^2\rangle>0$) when $h_w>h_{c}^{\text{Ising}}$, while the charge $\pmb{e}^2$ and flux $\pmb{m}^2$ do not condense ($\langle\pmb{e}^2\rangle=\langle\pmb{m}^2\rangle=0$).
Thus, the surviving anyons precisely match those of the DS topological order, see Sec.~\ref{app:Z4_anyon_cond}.
In the two-coupled quantum Ising model, 
the above phase transition is a conventional 0-form $\mathbb{Z}_2$ symmetry breaking transition,
where $\prod_i\sigma_i^z$ is spontaneously broken when $h_w>h_{c}^{\text{Ising}}$, so it belongs to the 3D Ising universality class.
In the perturbed $\mathbb{Z}_4$ QD model, the corresponding transition involves the breaking of a $\mathbb{Z}_2$ 1-form symmetry,
with $\prod_{e\in\hat{C}}W^2_e$ is spontaneously broken in the $\mathbb{Z}_4$ QD phase and preserved in the DS phase.
This is known as the Ising* transition~\cite{Universal_CFT_spec_2016_app}, which has the same critical exponents as the Ising transition.
Hence, in the critical regime, the condensation order parameter satisfies $\langle\pmb{e}^2\pmb{m}^2\rangle\propto (h_w-h_{c}^{\text{Ising}})^\beta$ and the area law coefficients satisfies $\alpha(\pmb{e})\propto \alpha(\pmb{m})\propto (h_w-h_{c}^{\text{Ising}})^{2\nu}$, where $\beta=0.326 418 (2)$ and $\nu=0.629970(4)$~\cite{Conformal_bootstrap_2016_app}.
As shown in Figs.~\ref{fig:ising}c and d, the double-log plots of the numerical condensation order parameters and the area law coefficients indicate that the numerical results are consistent with the theoretical analysis.

\begin{figure}[h]
    \centering
    \includegraphics[width=0.62\linewidth]{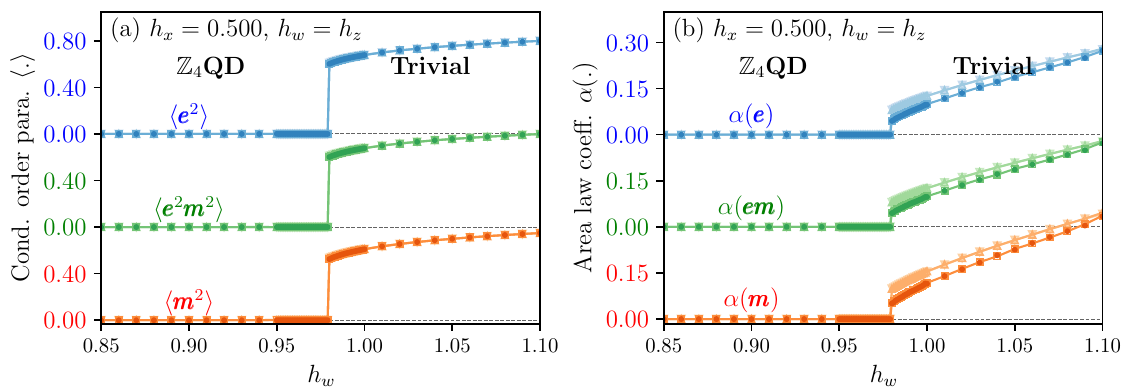}
    \caption{\textbf{Condensation order parameters and the area law coefficients along the line $h_x=-1.000$, $h_w=h_z$}.
A first-order transition occurs at $h_{w} = h_{z} \approx 0.980$.
Different colors correspond to different anyons, and the colors from light to dark correspond to the bond dimensions $(D,\chi)=(4,100),(4,160),(5,100),(6,100)$.
    (a) Condensation order parameters.
    (b) Area law coefficients.
    }
    \label{fig:1st_order}
\end{figure}

When $h_x =0.5$, there is no direct transition between the TC and DS phases, and the $\mathbb{Z}_4$ QD and trivial phases are separated by a first-order transition line, see Fig.~\ref{fig:phasediagrams_supp}a.
We perform simulations along the line where $h_w = h_z$ as an example, as illustrated in Fig.~\ref{fig:1st_order}.
Near the first-order transition point $(h_x, h_z, h_w) \approx (0.500, 0.980, 0.980)$, the condensation order parameters and the area law coefficients display discontinuities, which indicate a first-order phase transition.
Furthermore, we find that all expectation values converge as the iPEPS bond dimension $D \geq  5$, and confirm that these quantities remain unchanged with increasing CTMRG environment bond dimension $\chi$.

\section{Analysis the critical exponent at the TC-DS transition point using $O(2)$ field theory}\label{app:O2_phi_4_theory}
In this section, we show how to predict the critical exponents of the condensation order parameters and the area law coefficients at the TC-DS transition point, i.e. the multi-critical point where four phases meet.
A natural candidate for effective Ginzburg-Landau field theory at the multi-critical point of the two-coupled Ising model is the O(2) $\phi^4$ model described by the Lagrangian:   
\begin{equation}\label{eq:O(2)}
    \mathcal{L}=\left(\partial \pmb{\phi}\right)^2/2+m^2\pmb{\phi}^2/2+g\pmb{\phi}^4/(4!)+\cdots,
\end{equation}
where $m$ is the mass and $g$ is the coupling strength, and $\pmb{\phi}=(\phi_0,\phi_1)$ is a two-component order parameter field corresponding respectively to the two local order parameters $\langle \tau^z\sigma^x\rangle$ and $\langle \tau^z\rangle$. The Wilson-Fisher fixed point of the O(2) $\phi^4$ model belongs to the XY universality class~\cite{Wilson_fisher_1972_app,O_N_CFT_2023_app}. However, when the O(2) $\phi^4$ model serves as an effective theory for continuous topological phase transitions, people usually say the universality class is the XY*. Next, let us first discuss the detailed distinctions between XY and XY* theories and then we proceed to analysis of the critical exponents.

\subsection{Difference between XY and XY* theories}

The XY and XY* theory can be distinguished from the field theory perspective. In the XY* theory, $\phi_{0}$ can only be created in pairs, and $\phi_{0}$ and $-\phi_{0}$ are indistinguishable, so that both periodic and anti-periodic boundary conditions have to be considered. Namely the $\mathbb{Z}_2$ symmetry of $\phi_{0}$ field is gauged by orbifolding the $\phi_{0}$ field. Notice that the $\phi_{1}$ field remains unchanged. The example of the XY universality class is the transition from the paramagnetic phase to the complete $\mathbb{Z}_2\times\mathbb{Z}_2$ symmetry-breaking phase in the quantum Ashkin-Teller model or in our two-coupled Ising model. The example of the XY* universality class is the transition from the $\mathbb{Z}_2$ symmetry enriched topological phase to the trivial $\mathbb{Z}_2$ symmetry breaking phase in a $\mathbb{Z}_2$ toric code Ising model, which is obtained by gauging one of the $\mathbb{Z}_2$ symmetries of the quantum Ashkin-Teller model~\cite{Ising_toric_code_2023_app}. In summary, if the phase transition is a conventional Landau type, it corresponds to the XY universality class; if the phase transition involves symmetry-enriched topological phases with symmetry fractionalization, it is called the XY* universality class.

So, strictly speaking, the TC-DS transition is distinct from the XY* theory in Ref.~\cite{Ising_toric_code_2023_app} because the perturbed $\mathbb{Z}_4$ quantum double model is obtained by gauging \emph{both} two $\mathbb{Z}_2$ symmetries of the two-coupled Ising model [up to 3-cocycle twist, see Eq.~\eqref{eq:summary_transformation}], so there is no 0-form $\mathbb{Z}_2$ global symmetry and symmetry enriched topological phases. Therefore, when describing the TC-DS transition using the O(2) $\phi^4$ theory,  the $\mathbb{Z}_2$ symmetry of both $\phi_0$ and $\phi_1$ is gauged. In this work, we ignore the distinction between our multi-critical point and the symmetry-enriched topological phase transition in Ref.~\cite{Ising_toric_code_2023_app}, because roughly speaking, the universality class of conventional phase transitions is the non-starred version and the universality class of topological transitions is the starred version. For instance, people say the Ising transition for the transverse field Ising model and the Ising* transition for the $\mathbb{Z}_2$ toric code model in a field~\cite{Universal_CFT_spec_2016_app}.

We also would like to emphasize that although orbifolding $\phi_0$ and $\phi_1$ fields will project out the fields with odd parity, we can nevertheless find that their dual objects are the symmetry defects (anti-periodic boundary conditions), whose corresponding lattice operators are non-local semi-infinite string operators. A simple example is the periodic boundary critical Ising chain in the even sector, where the primary field $\sigma$ with the scaling dimension $1/8$ disappears, but its dual primary field $\mu$ with the scaling dimension $1/8$ appears in the even sector of the anti-periodic critical Ising chain~\cite{Topo_conformal_dfct_2016_app}, and the corresponding lattice operator is the non-local disorder parameter~\cite{Fradkin2017_app}. Therefore, if we do not restrict to local order parameters and take the non-local order parameters into consideration, it can be found that the local order parameters in the ungauged theory or model and the non-local order parameters in the gauged theory or model are in one-to-one correspondence. Next, we will discuss these order parameters and their critical exponents.

\subsection{Analysis the critical exponents}

\begin{table}[h]
\centering

\caption{Classifying fields in the O(2) $\phi^4$ model using irreps of the O(2), and the scaling dimensions of the fields are also shown.}

\begin{tabular}{C{3.0cm}C{2.0cm}C{4.0cm}}
\toprule
      field & irrep. & scaling dimension~\cite{O_2_exponent_2020_app} \\  
\midrule
$\phi^2_0+\phi^2_1$  & $\pmb{0}$  & 1.51136(22)  \\ 
$(\phi_0,\phi_1)$ & $\pmb{1}$ & 0.519088(22) \\  
$(\phi^2_0-\phi^2_1,\phi_0\phi_1)$ & $\pmb{2}$ & 1.23629(11) \\  
\bottomrule
\end{tabular}

\label{tab:primary_fields}
\end{table}

We can classify the fields in the O(2) $\phi^4$ model in terms of irreps of O(2) group, which includes the one-dimensional trivial irrep $\pmb{0}$, the one-dimensional sign irrep $\pmb{0}^{-}$ and two-dimensional irreps $\pmb{q}=q\oplus(-q)$, where the integer $q$ are the O(2) charges. Some fields transform under the irreps of O(2) and their scaling dimensions at the Wilson-Fisher fixed point are shown in Tab.~\ref{tab:primary_fields}.

Let us first consider the $\mathbb{Z}_4$ QD-trivial phase transition along the line (i): $h_w=h_z,~h_x=-1$. In the corresponding two-coupled Ising model, the lattice operator that derives this transition is $-\sum_{\langle i,j\rangle}\left(h_w\tau_i^z\tau_j^z\sigma_i^x\sigma_j^x+h_z\tau_i^z\tau_j^z\right)$, which is symmetric under the $\mathbb{Z}_2\times \mathbb{Z}_2$ symmetry. So, we expect that the corresponding field is $\phi_0^2+\phi_1^2$, because it is also invariant under the $\mathbb{Z}_2\times \mathbb{Z}_2$ symmetry: $\phi_{0}\leftrightarrow -\phi_{0}$ and $\phi_{1}\leftrightarrow -\phi_{1}$. 
Since the scaling dimension of $\phi_0^2+\phi_1^2$ is  $\Delta_{\pmb{0}}=1.51136(22)$, the correlation length critical exponent is $\nu_{\pmb{0}}=1/(3-\Delta_{\pmb{0}})=0.67175 (1)$. Because the quantity $\alpha[\cdot]\xi^2$ is dimensionless~\cite{xu_huang_2025_app}, the critical exponent of all area law coefficients is $2\nu_{\pmb{0}}$. Moreover, considering that the local order parameters $\langle \tau_i^z \rangle$ and $\langle \tau_i^z\sigma_i^x\rangle$ correspond to the fields $\phi_0$, $\phi_1$ with the scaling dimension $\Delta_{\pmb{1}}=0.519088(22)$ , we know $\langle\sigma_i^x\rangle$ corresponds to the field $\phi_{0}\phi_1$ with the scaling dimension $\Delta_{\pmb{2}}=1.23629(11)$. 
So the critical exponent of $\langle \tau_i^z \rangle$ and $\langle \tau_i^z\sigma_i^x\rangle$ is $\beta_{-}=\Delta_{\pmb{1}}\nu_{\pmb{0}}=0.34870(7)$, and the critical exponent of $\langle\sigma_i^x\rangle$ is $\beta_{\phi}=\Delta_{\pmb{2}}\nu_{\pmb{0}}=0.83048(2)$. 
Notice that the critical exponent $\eta$ of correlation functions can be obtained from scaling dimensions $\Delta$ via $\eta=2\Delta-d+2$, where $d=3$ is the spacetime dimension. From $\Delta_{\pmb{1}}$ and $\Delta_{\pmb{2}}$, we obtain $\eta_{\pmb{1}}\approx 0.04$ and $\eta_{\pmb{2}}\approx 1.47$, respectively. Previous work claims that in the XY theory $\eta=\eta_{\pmb{1}}$ and in the XY* theories $\eta=\eta_{\pmb{2}}$~\cite{XY_science_2012_app,Ising_toric_code_2023_app}. Actually, if we take both local and non-local order parameters into consideration, both $\eta_{\pmb{1}}$ and $\eta_{\pmb{2}}$ can be obtained from the XY or XY* theory. 
Mapping the local order parameters of the two-coupled quantum Ising model back to the condensation order parameters of the perturbed $\mathbb{Z}_4$ QD model, we know the critical exponents of $\langle \pmb{e}^2\rangle$ and $\langle \pmb{e}^2\pmb{m}^2\rangle$ are $\beta_{-}$, and the critical exponent of $\langle \pmb{m}^2\rangle$ is $\beta_{\phi}$.

Then, we consider the transition between the TC and DS phases along the line (ii): $h_x=-1,~h_w=2h_c-h_z$. When perturbing away from the critical point along this line, the $\mathbb{Z}_2$ symmetry exchanging $h_z$ and $h_w$ is broken. This corresponds to breaking the $\mathbb{Z}_2$ symmetry $\phi_0\leftrightarrow \phi_1$ in the field theory. Thus, the field  $\phi_0^2-\phi_1^2$ with the scaling dimension is $\Delta_{\pmb{2}}=1.23629(11)$ is a candidate relevant field perturbing from the multi-critical point towards TC or DS phases. Therefore, the correlation length critical exponent $\nu_{\pmb{2}}=1/(3-\Delta_{\pmb{2}})=0.56699 (4)$ determining the critical exponent of the area law coefficients is different from that of the transition between the $\mathbb{Z}_4$ QD and the trivial phases. Because in the two-coupled quantum Ising model, $\langle \tau_i^z\sigma_i^x\rangle$ and $\langle\tau_i^z\rangle$ correspond to $\phi_0$  and $\phi_1$ with the scaling dimension $\Delta_{\pmb{1}}=0.519088(22)$, the critical exponent of $\langle \tau_i^z\sigma_i^x\rangle$ is $\beta_{\phi}'=\Delta_{\pmb{1}}\nu_{\pmb{2}}=0.519088(22)\times0.56699 (4)=0.29432(3)$.  Mapping the local order parameters of the two-coupled Ising model back to the condensation order parameters of the perturbed $\mathbb{Z}_4$ QD model, we know the critical exponents of $\langle \pmb{e}^2\rangle$ and $\langle \pmb{e}^2\pmb{m}^2\rangle$ are $\beta'_{\phi}$.

Next, we discuss the stability of the continuous TC-DS transition under varying $h_x$. The term $h_x\sum_{\langle i,j\rangle}\sigma^x_i\sigma^x_j$ is invariant under the $\mathbb{Z}_2\times\mathbb{Z}_2$ spin-flip symmetry and the $\mathbb{Z}_2$ symmetry exchanging $h_z$ and $h_w$. Considering $\sigma^x_i$ corresponds to the field $\phi_0^2\phi_1^2$~\cite{Ising_toric_code_2023_app}, we expect that $\sigma^x_i\sigma^x_j$ corresponds to the field $\phi_0^2\phi_1^2$, which is a basis of the O(2) irrep $\pmb{4}$ with a scaling dimension $3.14(2)>3$~\cite{O_2_exponent_2020_app}, so it is an irrelevant perturbation and the continuous TC-DS transition point is stable under varying $h_x$. 

Finally, since it is also proposed that the O(2) $\phi^4$ theory can describe the topological multi-critical point of the standard $\mathbb{Z}_2$ TC model in a tilted field~\cite{MCP_TC_2022_app}, let us compare the multi-critical point of the $\mathbb{Z}_2$ TC model in a tilted field and that of the perturbed $\mathbb{Z}_4$ QD model. 
The $\mathbb{Z}_2$ TC model in a tilted field has phase transitions from the TC phase to the trivial phase by condensing either $\mathbb{Z}_2$ charge excitations or $\mathbb{Z}_2$ flux excitations. Along the self-dual line, the duality symmetry swaps the $\mathbb{Z}_2$ charge and flux excitations, and there is a topological multi-critical point separating the TC phase and a first-order line where the duality symmetry is broken spontaneously~\cite{TC_phase_diagram_expansion_2009_app,MCP_TC_2010_app,Youjin_2012_app}.
Although it is proposed that the O(2) $\phi^4$ model can describe the topological multi-critical point of the  $\mathbb{Z}_2$ TC model in a tilted field~\cite{MCP_TC_2022_app}, the non-trivial mutual statistics between the $\mathbb{Z}_2$ charge and the flux excitations of the TC model are not included in the O(2) $\phi^4$ model because the two fields $\phi_0$ and $\phi_1$ commute with each other. Therefore, it is argued that the multi-critical point of the standard TC might not be the XY* universality class~\cite{Adam_Nahum_2021_app}. In contrast, there is no such problem at the multi-critical point of the perturbed $\mathbb{Z}_4$ QD model, because the mutual statistics between $\pmb{e}^2$ and $\pmb{e}^2\pmb{m}^2$ are bosonic.

\end{widetext}
\end{document}